\documentclass[superscriptaddress,preprintnumbers,amsmath,amssymb,nofootinbib, prd, twocolumn]{revtex4}
\usepackage{amssymb}
\usepackage{amsmath}
\usepackage{hyperref}
\usepackage{color}
\usepackage{graphicx}
\usepackage{soul}
\newcommand{\be}{\begin{equation}}
\newcommand{\ee}{\end{equation}}
\newcommand{\bea}{\begin{eqnarray}}
\newcommand{\eea}{\end{eqnarray}}
\newcommand{\tr}{{\rm Tr}}

\newcommand{\Eqref}{Eq.\eqref}

\begin{document}

\title{Continuum limit in matrix models for quantum gravity from the Functional Renormalization Group}
\author{Astrid Eichhorn}
\affiliation{\mbox{\it Perimeter Institute for Theoretical Physics, 31 Caroline Street N, Waterloo, Ontario, N2L 2Y5, Canada}
\mbox{\it E-mail: {aeichhorn@perimeterinstitute.ca}}}

\author{Tim Koslowski}
\affiliation{\mbox{\it Department of Mathematics and Statistics, University of New Brunswick, Fredericton, New Brunswick E3B 5A3, Canada}
\mbox{\it E-mail: {t.a.koslowski@gmail.com}}}

\begin{abstract}
We consider the double-scaling limit in matrix models for two-dimensional quantum gravity, and establish the nonperturbative functional Renormalization Group as a novel  technique to compute the corresponding interacting fixed point of the Renormalization Group flow. We explicitly evaluate critical exponents and compare to the exact results. The functional Renormalization Group method allows a generalization to tensor models for higher-dimensional quantum gravity and to group field theories. As a simple example how this method works for such models, we compute the leading-order beta function for a colored matrix model that is inspired by recent developments in tensor models.
\end{abstract}

\maketitle

\section{Introduction}
In path-integral approaches to quantum gravity, the sum over geometric (and topological) configurations can be tackled  by the  introduction of an (unphysical) discretization in order to explicitly construct all configurations that contribute to the expectation value of a given quantity. The partition function then becomes a sum over discretized random surfaces, either of fixed topology, or possibly even involving a summation over topologies.
Quantum gravity corresponds to the continuum limit, at which the discretization scale is taken to zero. This idea underlies Causal and Euclidean Dynamical Triangulations \cite{Ambjorn:1991pq, Ambjorn:2012jv,Ambjorn:2011cg}, as well as the more recently developed tensor models \cite{Rivasseau:2011hm, Rivasseau:2012yp} and group field theories \cite{Boulatov:1992vp,Freidel:2005qe,Oriti:2007qd,Oriti:2011jm}. The first successful implementation of this idea goes back to the partition function for two-dimensional Euclidean quantum gravity, which includes an integral over two-dimensional geometries and a sum over topologies:
\be
Z \sim \sum_{\rm topologies} \int \mathcal{D}g_{\mu \nu}\, e^{-\beta A + \gamma \chi},\label{Z2d}
\ee
where $g_{\mu \nu}$ is the metric and $A= \int \sqrt{g}$ is the surface area. $\beta$ denotes the cosmological constant  and $\gamma$ is related to the Newton coupling. $\chi = \frac{1}{4 \pi} \int \sqrt{g} R = 2- 2h$ is the Euler character and $h$ denotes the number of handles of the surface. The quantum theory is nontrivial, because the partition function involves a sum over topologies, despite the fact that the scalar curvature term $\int \sqrt{g}R$ is topological, and therefore does not contribute to the equations of motion.
 We can approximate the surfaces that are being summed over by discrete triangulations, or more general "polygonizations", so that $Z$ corresponds to the sum over random polygonizations of different topologies. The integral over the geometry was first treated as a sum over discretized randomly triangulated surfaces in \cite{Weingarten:1982mg}.
 In a next step, we use a matrix integral as the generating functional for the random triangulations, which allows us to take the continuum limit of the sum over triangulations. This step makes use of the fact that the dual to a Feynman graph of the matrix model  with trivalent vertices corresponds to a random triangulation. Using $ N\times N$  hermitian matrices $\phi$, the partition function $Z_N$ generates Feynman diagrams which are dual to triangulations, as we are using a tri-valent vertex $\sim \phi^3$:
\be
Z_N = \int d\phi\, e^{-\frac{1}{2} \tr \phi^2 + \frac{g}{\sqrt{N}} \tr \phi^3} 
\ee
This model automatically has a $U(N)$ symmetry. Alternatively, we could use a $\phi^4$ interaction corresponding to a four-valent vertex and thus a "squarulation" of the surface. In that case, the model would exhibit an additional $Z_2$ symmetry $\phi \rightarrow - \phi$.
Since both connected as well as non-connected triangulations are generated by the matrix model partition function, the generating functional of connected correlators, i.e., the free energy, $N^2 Z(N,g) = \ln Z_N$, will correspond to the 2-d gravity partition function $Z$ in the continuum limit. The continuum limit is reached by taking $g\rightarrow g_c$. $\langle A \rangle$ diverges at this critical value of the coupling. Thus the area of the individual triangles can be taken to zero to give a continuum surface with finite area. The correspondence between $g \rightarrow g_c$ and the continuum limit can also be understood, as the perturbation series diverges in this limit. Accordingly the generating functional is dominated by diagrams with a diverging number of vertices, which implies that the dual triangulations approach a continuum surface. The large $N$ expansion allows one to sort the contributions by the genus of the triangulated surface, as the generating functional can be decomposed into the 
contributions from surfaces with genus $h$ as follows
\be
Z= \sum_h N^{2-2h} Z_h.
\ee

In the limit $N \rightarrow \infty$, the limit $g \rightarrow g_c$ thus only takes into account planar (i.e., spherical) surfaces. 
To retain the contribution from higher-genus surfaces, the limit $N \rightarrow \infty$ and $g \rightarrow g_c$ has to be taken simultaneously: As each $Z_h$ diverges as $g \rightarrow g_c$, the $N^{-2h}$ suppression can be compensated in this limit, as 
\be
Z_h \sim (g_c -g)^{(2- \gamma_{\rm str})(2-2h)/2},
\ee
where $\gamma_{\rm str}$ is a critical exponent. Continuum Liouville theory predicts $\gamma_{\rm str} = -1/2$ for the case of pure gravity \cite{Di Francesco:1993nw}. 

Accordingly, the so-called double-scaling limit, in which $g \rightarrow g_c$ and $N \rightarrow \infty$ while one fixes
\be
N \left(g- g_c \right)^{(2-\gamma_{\rm str})/2} = \rm const,\label{doublescaling}
\ee
retains contributions from all surfaces with higher genus to the generating functional \cite{Douglas:1989ve,Brezin:1990rb,Gross:1989vs}. In this limit, the average number of building blocks in triangulations of any genus $h$ diverges, while the "lattice spacing" of the random discretized surface is taken to zero. 
For reviews see \cite{Ginsparg:1993is, Di Francesco:1993nw,Ambjorn:1994yv,Marino:2004eq}.
For this work, it will be crucial that \Eqref{doublescaling} can be understood as a particular scaling of $g$ with $N$ in the vicinity of a critical point $g_c$. In other words, the double scaling limit can be understood as a fixed point of the Renormalization Group flow with $N$. 

We follow \cite{Brezin:1992yc} and assume that the partition function satisfies a Callan-Symanzik-type equation
\be
\left( N \frac{\partial}{\partial N}+ \beta_g \frac{\partial}{\partial g} + \gamma(g)\right)Z(N,g) = r(g),
\ee
where $\beta_g = N \partial_N g(N)$.
At a fixed point $g= g_{\ast}$, where $\beta_g\vert_{g= g_{\ast}}= 0$, the singular part of the partition function will accordingly satisfy a scaling relation of the form
\be
Z(N,g) = \left( g_c-g \right)^{\gamma_1} f \left((g_c-g) N^{\frac{2}{\gamma_1}} \right).
\ee
Here we have to identify 
\be
\gamma_1 = \frac{2}{\beta'(g_{\ast})}, 
\ee
and $\gamma_{\rm str} = 2- \gamma_1$. Accordingly the pure gravity case requires $\beta'(g_{\ast}) = 4/5$. For the generalization to an effective action containing further operators, $\beta'(g_{\ast})$ has to be replaced by the relevant critical exponent and the corresponding fixed point can only exhibit one relevant direction. 

In this paper we will set up a Renormalization Group flow in the Wilsonian sense, integrating out matrix entries by going from $N \rightarrow N+ \delta N$. As in the standard effective field theory setting, this process will generate further operators in the effective action for the low- $N$ degrees of freedom. The $U(N)$ symmetry of the $N \times N$ hermitian matrix model will restrict the new operators to be of the form $\tr \phi^i_1\, \cdot \dots \cdot \tr \phi^{i_n}$.

Note that the double scaling limit corresponds to an interacting fixed point $g_{\ast} \neq 0$. The fact that only one coupling needs to be tuned corresponds to this fixed point only having one ultraviolet-relevant direction. 

This approach has mainly been developed to understand string theory coupled to matter with central charge $c>1$, see, e.g., \cite{Alfaro:1992nq,Higuchi:1993pu,Higuchi:1994rv,Dasgupta:2003kk}. Here, our motivation comes from another direction, namely the generalization of matrix models to higher-dimensional models of discrete random surfaces, i.e., non-string theory models for $d=3,4$ dimensional quantum gravity.
We are interested in developing a technique that can be generalized to the type of tensor models that is currently investigated in the context of quantum gravity models in more than two dimensions. The analogy to matrix models is as follows: In the same way in which matrix models generate triangulations of 2-dimensional surfaces, models using 3-rank or 4-rank tensors yield  3 or 4 dimensional simplicial geometries, respectively, \cite{Ambjorn:1990ge,Sasakura:1990fs,Gross:1991hx}. In these models, the continuum limit again corresponds to the large-$N$ limit, and a large $N$ expansion exists for the class of colored tensor models \cite{Gurau:2009tw, Gurau:2010ba,Gurau:2011aq,Gurau:2011xq}.
The double-scaling limit in such models is subject of current research \cite{Gurau:2011sk,Dartois:2013sra}.  The method that we introduce in this paper will allow one to investigate whether a double-scaling limit exists and what its universality class is. As we will show using a colored matrix model, our method is extendible to the case of colored tensor models, which are of particular interest, as they triangulate less singular spaces \cite{Gurau:2010nd} and admit a $1/N$ expansion.
 
These colored tensor models are closely related to group field theory \cite{Freidel:2005qe,Oriti:2007qd,Oriti:2011jm}. There, a quantum field theory is constructed where the field is a map from a group manifold to the real (or complex) numbers, thus incorporating the spin-foam amplitudes of covariant loop quantum gravity into a tensor model. Using the generalization of a Fourier transform on the group manifold, these models can be mapped to tensor models. Here, the physical interpretation of the dual to a Feynman graph is again a triangulation. The theory can then have (at least) two interesting phases: One corresponds to a pregeometric phase in terms of disconnected building blocks (similar to a gaseous phase). A second phase corresponds to a condensation of these fundamental building blocks into a continuous spacetime. For a second-order phase transition between these phases, the universal critical exponents correspond to those arising at a fixed point of the Renormalization Group. Thus the search for (non)
-interacting fixed points in these models is of interest to determine whether such phase transitions occur. Such a  phase transition can then either be interpreted as a point to take the continuum limit, as in the case of CDTs \cite{Ambjorn:2011cg}, or as a physical phase transition, in which case the building blocks are interpreted as fundamental physical building blocks, in constrast to being a regularization as in the first case.

This paper is structured as follows: In sec. \ref{hmm} we review the Renormalization-Group setup put forward in \cite{Brezin:1992yc} that consists in integrating out matrix entries explicitly. We then introduce the functional Renormalization Group and the Wetterich equation for matrix models. We show how to define the canonical scaling dimensionality of the couplings with $N$, and then derive nonperturbative $\beta$ functions. We prove an equation for the $\beta$ functions in a truncation involving single-trace terms to arbitrarily high order.
We show that our novel method employing the functional Renormalization Group can reproduce the results obtained in \cite{Brezin:1992yc, Ayala:1993fj}. We also demonstrate how to include multi-trace operators in the Renormalization Group flow. We then show how to treat a colored matrix model, and derive $\beta$ functions in a simple approximation in sec.~\ref{cmm}. Before concluding in sec.~\ref{conclusion}, we describe the general recipe how to apply the functional Renormalization Group in matrix and tensor models in sec.\ref{recipe}.

\section{Hermitian Matrix Model}\label{hmm}
\subsection{Explicit integration over matrix entries}
In the following, we will focus on the matrix model with a $\phi^4$ interaction, corresponding to a "squarulation" of the surface. The continuum limit is universal and independent of the choice of polygons used to construct the discrete simplicial geometry. The additional $Z_2$ symmetry of this model restricts the operators that are generated by the Renormalization Group flow.
To derive the $\beta$ functions for the hermitian matrix model, it has been suggested  in \cite{Brezin:1992yc}  to explicitly integrate out the $(N+1)$th row and column in the path-integral over $(N+1)\times (N+1)$ matrices. Let us briefly review the technique and the results of  \cite{Brezin:1992yc, Ayala:1993fj}, see also \cite{Higuchi:1993pu,Higuchi:1994rv}.
We consider the action
\be
S_{N+1}[\phi_{N+1}] = (N+1) \tr \left(\frac{\phi_{N+1}^2}{2}+ \frac{g_4}{4}  \phi_{N+1}^4+ \frac{g_6}{6} \phi_{N+1}^6\right),
\ee
which has a $Z_2$ symmetry under $\phi_{N+1} \rightarrow - \phi_{N+1}$.
We then parameterize the matrix $\phi_{N+1}$ by the matrix $\phi_N$ and a complex $N$-component vector $v$ for the $(N+1)$th column and its complex conjugate $v^{\ast}$ for the $(N+1)$th row. The last diagonal entry in $\phi_{N+1}$ is denoted by a complex number $\alpha$. We now explicitly integrate out the vector $v$ in the path integral, which yields an effective action for $\phi_N$. In this step, we will only perform the Gau\ss{}ian integral over $v$, and neglect $\alpha$, which is appropriate in the large-$N$ limit.
Re-exponentiating the determinant, and expanding the logarithm, we obtain
\bea
S'[\phi_N] &=& (N+1) \tr \left(\frac{1}{2}\phi_N^2+\frac{g_4}{4} \phi_N^4 + \frac{g_6}{6} \phi_N^6\right)\nonumber\\
&+& g\, \tr\left( \phi_N^2\right) - \frac{g_4^2}{2} \tr \left(\phi_N^4\right)+g_6 \tr \left(\phi_N^4\right).
\eea
In order to recover the canonical normalization of the quadratic term in $\phi_N$, we perform a rescaling
\be
\phi_N = \rho \phi_N',
\ee
with 
\be
\rho = 1- \frac{2g_4+1}{2N} + \mathcal{O}\left(\frac{1}{N^2} \right).
\ee
It is then straightforward to read off the relation between the couplings $g_4'$ and $g_4$
\be
\frac{N\, g_4'}{4}= (N+1) \frac{g_4}{4} \rho^4.
\ee 
With the definition
\be
g_4-g_4'= \frac{1}{N} \beta(g)
\ee
we then obtain the following $\beta$ functions, where terms $\mathcal{O} \left(\frac{1}{N} \right)$ are neglected:
\bea
\beta_{g_4} &=& g_4+6g_4^2 +4 g_6\label{betagZJ},\nonumber\\
\beta_{g_6}&=& -g_6\left(1-3(1+2g_4) \right)+ 6 g_4 g_6 - 6g_4^3.
\eea
For $g_6=0$ this is the result obtained in \cite{Brezin:1992yc} (note that our definition of the $\beta$ function differs by a sign) and extended to higher order in the couplings and to include further couplings in \cite{Ayala:1993fj}. We obtain
\be
g_{4\, \ast}= -0.1057, \quad g_{6\, \ast}= -0.0097.
\ee
For the critical exponents\footnote{These determine whether a coupling must be tuned in order to reach a particular fixed point: From \Eqref{thetadef} we can deduce the solution to the linearized flow around the fixed point 
\be
\beta_{g_i}(\{g_n\})= \sum_j\frac{\partial \beta_{g_i}}{\partial g_j} \Big|_{g_n = g_{n\, \ast}}\left(g_j - g_{j\, \ast} \right) +\mathcal O((g_j-g_{j\,\ast})^2)\,.\label{linflow}
\ee
The solution to this linearized equation reads
\be
g_{i}(N) = g_{i\, \ast} + \sum_I C_I V^{I}_i  \left(\frac{N}{N_0}\right)^{- \theta_I},
\ee
where 
\be
- \frac{\partial \beta_{g_i}}{\partial g_j} \Big|_{g_n = g_{n\, \ast}} V_I = \theta_I V_I.
\ee
Herein, $C_I$ is a constant of integration and $N_0$ is a reference "scale". The $V_I$ are the eigenvectors and $-\theta_I$ the eigenvalues of the stability matrix,  which is defined by \Eqref{linflow}. The additional negative sign in $\theta_I$ is useful, because then $\theta_I$ equals the canonical dimensionality at a non-interacting fixed point.
In order to approach the fixed point in the IR, observe that the $C_I$ are arbitrary for irrelevant directions where $\theta_I <0$. In contrast, a relevant direction with $\theta_I>0$ corresponds to a parameter that needs to be tuned in order to ensure that the fixed point is reached in the IR.}, generally defined by
\be
\theta_i = - {\rm eig} \left(\frac{\partial \beta_{g_i}}{\partial g_j} \right)\Big|_{g_k = g_{k\, \ast}},\label{thetadef}
\ee
we obtain
\be
\theta_1=1, \quad \theta_2=-1.46. \label{critexpZJ}
\ee
Let us discuss the significance of this result: As expected, the fixed point has one relevant direction, as only $g_4$ needs to be tuned in the double-scaling limit. We observe that $g_6$, although itself an irrelevant coupling, yields a significant contribution to $\beta_{g_4}$, and improves $g_{4\, \ast}$ from the value $-1/6$ which is the fixed-point value for $g_6=0$.

The exact value $g_c = -1/12$ is approximated by our result, and at a first glance, the value of $\theta$ does not seem to be too different from the exact value $\theta = -0.8$. In fact, this value can be traced to the "canonical dimensionality" of the  coupling, which yields the term $\sim g$ in the $\beta$ function: In order to obtain a nontrivial fixed point in 
\be
\beta_g=g(1+c \,g),
\ee
where $c$ is a constant, $g_{\ast}= -\frac{1}{c}$. Then, 
\be
\theta=- \partial_g \beta_g \vert_{g= g_{\ast}}= -(1+c \, g_{\ast}) - g_{\ast} c = 1.
\ee
Accordingly we observe that at this order, the value of the critical exponent is determined by the dimensionality, and does not depend on the fixed-point value at all.

In the following, we will introduce a method that generalizes the idea to integrate out matrix entries, such that it applies beyond the perturbative regime, and can be generalized directly to higher-dimensional tensor models.
We will set up a nonperturbative Renormalization Group equation that allows to determine the scaling of couplings with the matrix size $N$.

To that end, let us first introduce the notion of scaling dimensionality in these models.

\subsection{Canonical scaling dimensionality}
As this model does not have a natural notion of momentum -- it can be viewed as a zero-dimensional field theory in this sense -- there is no standard canonical dimensionality of the couplings. Spacetime, and the attribution of momentum dimensions as familiar from high-energy physics, only emerges in the continuum limit. Nevertheless, couplings have an inherent dimensionality that determines their behavior under rescalings in $N$, just as the canonical dimensionality in standard quantum field theories determines the canonical scaling under standard scale-transformations. From the expression \Eqref{betagZJ}, one can distinguish two types of contributions to the $\beta$ function: There are contributions proportional to one power of the coupling itself, or higher powers of the coupling and higher-order couplings. The same structure appears in $\beta$ functions in high-energy physics. There it is clear that a loop-diagram that yields the contribution of quantum fluctuations to the $\beta$ function of $g_i$ cannot 
be proportional to one power of $g_i$ only. Instead it arises from the 
canonical dimensionality (and is therefore absent, e.g., in QCD in $d=4$.) In the same way, this type of contribution reflects the scaling dimensionality of the matrix model couplings under a rescaling in $N$.

The canonical scaling dimensionality of the couplings is determined in the following way: Considering only single-trace terms, the action can be written as
\begin{equation}
\bar{S}= \frac{Z_{\phi}}{2} \tr \left(\phi^2\right) +\sum_i \bar{g}_i \tr \left(\phi^i\right), \label{action}
\end{equation}
where $Z_{\phi}$ is a wavefunction renormalization. 
In order to establish the contact to the generating functional for 2d quantum gravity, a rescaling of the fields $\phi  \rightarrow \sqrt{N} \phi$ and couplings is useful, such that
\begin{equation}
S= N\left(\frac{Z_{\phi}}{2} \tr \left(\phi^2\right) +\sum_i \tilde{g}_i \tr \left(\phi^i\right) \right),
\end{equation}
as then each Feynmandiagram is weighted by a factor $N^{\chi}$, where $\chi$ is the Euler character of the surface associated to the diagram. This holds, as in the above parameterization, each vertex contributes a factor $N$, each propagator a $N^{-1}$ and each face, i.e., each closed loop, $N$ due to the associated index summation. Thus each diagram contributes with $N^{V-E+F}$, where $V, E$ and $F$ count the number of vertices, edges and faces, respectively. As $V-E+F = \chi$, the identification $N = e^{\gamma}$ then allows to make contact with the generating functional for 2d quantum gravity, cf. \Eqref{Z2d}. We therefore read off the following relation between $\tilde{g}_i$ and $\bar{g}_i$:
\begin{equation}
\bar{g}_i  N^{\frac{i-2}{2}}= \tilde{g}_i.
\end{equation}
This analysis agrees with the corresponding terms in the perturbative $\beta$ function in \cite{Brezin:1992yc, Ayala:1993fj}, cf. \Eqref{betagZJ}.

We observe that even the coupling with the lowest number of fields, $g_4$, is power-counting irrelevant, as it has dimensionality $-1$. We conclude that quantum fluctuations will be responsible for a shift towards relevance at the fixed point corresponding to the double-scaling limit.

To assign canonical dimensionality to terms with several traces, such as $\bar{g}_{i, j} \tr \phi^i\, \tr \phi^j$, we demand that each additional trace should be suppressed by a factor $\frac{1}{N}$. This assignment cancels the additional factor of $N$ that is associated with the additional trace.

Accordingly this yields
\be
\bar{g}_{i, j} N^{\frac{i+j}{2}} = \tilde{g}_{i, j}.\label{doubletracescaling}
\ee
At the same order in $\phi$, a double-trace term accordingly is shifted by one into irrelevance in comparison to the corresponding single-trace operator. This is consistent with the expectation that only one coupling should be tuned in the double-scaling limit. For the single trace operators, the coupling $g_4$ of dimensionality $-1$ is shifted into relevance, but not the coupling $g_6$ of dimensionality $-2$, cf. \Eqref{critexpZJ}. The assignment of dimensionality $-2$ to the first two-trace-operator $(\tr \phi^2)^2$ already suggests that it will not be shifted into relevance, if we assume that the contribution of quantum fluctuations to the scaling of different operators is of similar size.

If we further rescale the kinetic term to its canonical coefficient $1/2$, the renormalized dimensionless couplings $g_i$ are then given by
\be
g_i= \frac{\tilde{g}_i}{Z_{\phi}^{i/2}} = \frac{\bar{g}_i N^{\frac{i-2}{2}}}{Z_{\phi}^{i/2}}, \quad g_{i,j}= \frac{\bar{g}_{i, j} N^{\frac{i+j}{2}}}{Z_{\phi}^{\frac{i+j}{2}}}.\label{canondim}
\ee

Accordingly, the $\beta$ functions will take the following form
\be
\beta_{g_i}= \frac{i-2}{2}g_i + \frac{i}{2} \eta g_i +\dots,\label{equ:BetaType}
\ee
where $\eta = - \partial_t \ln Z_{\phi}$ and further terms are induced by quantum fluctuations. This form agrees with the $\beta$ functions that can be derived by explicit integration over matrix entries, cf. \cite{Brezin:1992yc, Ayala:1993fj}.
Let us now introduce a novel tool to perform this integration and derive the nonperturbative form of the $\beta$ functions.

At this point is is useful to comment on the universality of the fixed-point values: As is well-known, for a dimensionless coupling, the one-loop and two-loop coefficients of the $\beta$ function are universal, but higher orders are not. For dimensionful couplings, such as all couplings in our case, even the one-loop coefficient is nonuniversal. This implies that fixed-point values will not be universal, either. In contrast, the critical exponents are universal. Accordingly a comparison of $g_{\ast}$ to $g_c$ is less meaningful than the comparison of the critical exponents with the exact result.

\subsection{Set-up of the flow equation}
To generalize the perturbative setting of integrating over matrix entries, we use the framework of the nonperturbative functional Renormalization Group, \cite{Wetterich:1993yh}, for reviews see  \cite{Berges:2000ew,Polonyi:2001se,Pawlowski:2005xe,Gies:2006wv,Delamotte:2007pf,Rosten:2010vm,metzner2011,Braun:2011pp}.

Our approach here is closely related to similar developments in \cite{Sfondrini:2010zm}, where the functional Renormalization Group was used to demonstrate the property of asymptotic safety in the Grosse-Wulkenhaar model \cite{Grosse:2004yu, Disertori:2006nq}. A main difference to the present setting is the absence of a nontrivial kinetic term in our case, and the implications for the canonical dimensionality and the implementation of the matrix cutoff in the following. Note that this approach crucially differs from former applications of the functional Renormalization Group to matrix models in the context of condensed-matter models, see, e.g., \cite{Berges:2000ew}: There, quantum fluctuations of matrix-valued fields that are function on spacetime, are integrated out according to their momentum.

To derive the change of the generating functional under an integration of matrix entries between $N$ and $N + \delta N$, we follow the usual steps in setting up a flowing effective action: We consider the generating functional
$\mathcal{Z} = \int_{\Lambda} d \phi e^{-S[\phi]+ J \cdot \phi}$,
where $\Lambda$ is a UV cutoff on $N$ and $J$ is a source. We then define an $N$-dependent generating functional, where $N$ will be an infrared cutoff on the matrix size, i.e., we will integrate out matrix entries above $N$, but not below. 
We introduce a "mass-like" regulator function 
\be
\Delta S_N [\phi]=\frac{1}{2} \phi_{ab} R_N(a,b)_{ab\, cd} \phi_{cd},\label{DeltaS}
\ee
 where we now explicitly indicate the matrix indices $a,b,\dots$ for clarity. The function $R_N(a,b)$ can have any form compatible with the following three requirements:
\bea
\underset{a/N\rightarrow0, c/N\rightarrow0} {\rm lim} R_N(a,b)_{ab\, cd}  &>&0,\label{IRsup}\\
\underset{N/a\rightarrow0, N/a \rightarrow 0}{\rm lim} R_N(a,b)_{ab\, cd} &=&0\label{IRlim}\\
\underset{N\rightarrow \Lambda \rightarrow \infty} {\rm lim}R_N(a,b)_{ab\, cd} &\rightarrow&\infty.\label{UVlim}
\eea
Introducing this mass-type term into the generating functional
\be
\mathcal{Z}_N = \int_{\Lambda} d \phi\, e^{-S[\phi] - \Delta S_N[\phi]+J \cdot \phi},
\ee
yields a suppression of the matrix entries in the block $a,b =1, \dots, N$, cf. \Eqref{IRsup}. In contrast, the "UV" matrix entries with indices $a,b>N$ are integrated out. Thus the generating functional $\mathcal{Z}_N$ contains the effect of "UV" quantum fluctuations above $N$. Taking the limit $N \rightarrow 0$, the suppression term should approach zero, such that all matrix entries are integrated out in this limit, cf. \Eqref{IRlim}. This is a direct generalization of the idea put forward in \cite{Brezin:1992yc} to integrate the $(N+1)$th row and column of the matrices explicitly.

We now define an $N$ dependent flowing action by a modified Legendre transform
\be
\Gamma_N[\varphi] = \underset{J}{\rm{sup}} \left( J \varphi - \ln \mathcal{Z}_N\right)- \Delta S_N[\varphi]. 
\ee
The divergence of $\Delta S_N$ for $N\rightarrow \Lambda \rightarrow \infty$, cf. \Eqref{UVlim}, ensures that $\Gamma_{N \rightarrow \Lambda} \simeq S_{\rm bare}$, as it enforces a saddle-point evaluation of the path-integral. For $N \rightarrow 0$, $\Gamma_N \rightarrow \Gamma$, which is the full effective action and contains the effect of all quantum fluctuations.

\begin{figure}[!here]
\includegraphics[width=0.4\linewidth]{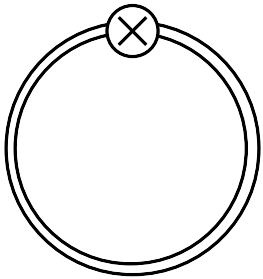}
\caption{\label{oneloop} The two matrix indices generate a "ribbon graph" with double lines. The crossed circle denotes the regulator insertion $\partial_t R_N$, while the double line denotes the full, field dependent propagator.}
\end{figure}

It is now straightforward to derive an equation encoding the change of $\Gamma_N$ under a change of $N$, which is the Wetterich equation for the matrix model case:
\be
\partial_t \Gamma_N = \frac{1}{2} {\textrm{tr}} \left(\Gamma_N^{(2)}+ R_N \right)^{-1} \partial_t R_N,
\ee
where $t = \ln N$. $\Gamma_N^{(2)}$ denotes the second functional derivative with respect to the degrees of freedom, in this case the entries of matrices.  $\left(\Gamma_N^{(2)}+ R_N \right)^{-1}$ is nothing but the regularized, nonperturbative field-dependent propagator of the theory. Accordingly the operator trace $\textrm{tr}$ in the Wetterich equation yields a one-loop form, cf. fig.~\ref{oneloop}. It is important to realize that the derivation of the Wetterich equation does not rely on the existence of a small parameter. It is therefore applicable in the nonperturbative regime. As a technical advantage, it takes a one-loop form, while the presence of the full propagator includes nonperturbative effects in this setting. 

\begin{figure}[!here]
\includegraphics[width=0.25\linewidth]{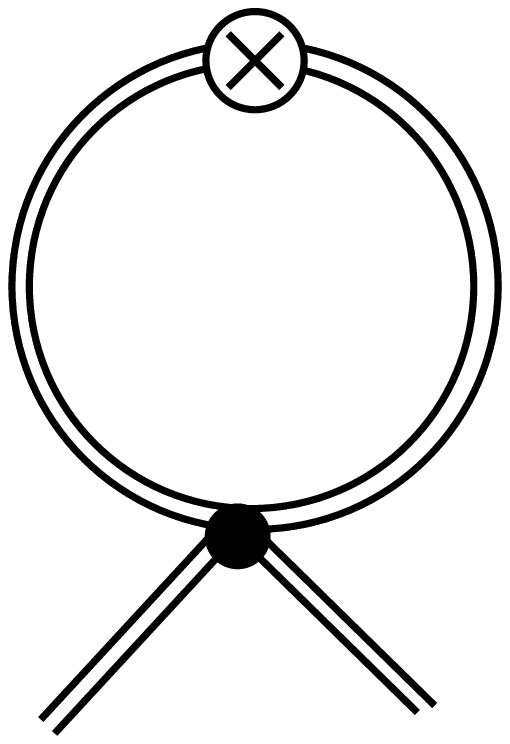} \quad \includegraphics[width=0.55\linewidth]{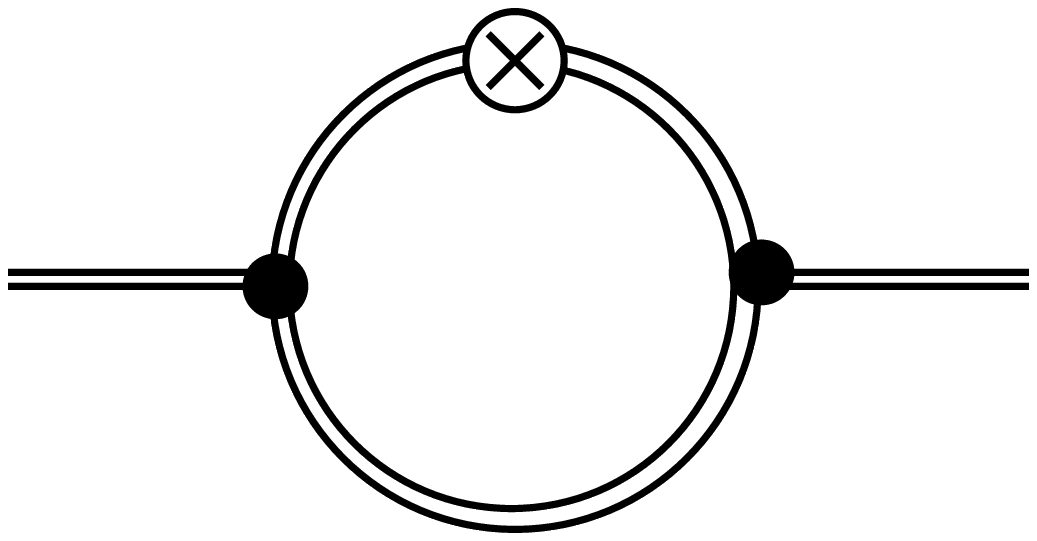}
\caption{\label{wavefunction} The derivative of the Wetterich equation generates vertices with a coupling to external fields. As an example, we show the effect of the first two derivatives, which generate a tadpole diagram and a two-vertex diagram, both contributing to the flow of the wave-function renormalization. For the $Z_2$ symmetric matrix model that we consider here, where $\tr \phi^3$ is not an operator in the action, the two-vertex diagram vanishes.}
\end{figure}

In order to obtain the $\beta$ function of a particular coupling, it is necessary to project the flowing action onto that coupling, by taking appropriate derivatives with respect to $\phi$. If the same projection prescription is applied to the  Wetterich equation, this yields the $\beta$ function of the corresponding coupling. Diagrammatically, the derivatives with respect to $\phi$ generate vertices with a coupling to an external $\phi$, as $\frac{\partial}{\partial \phi} \left(\Gamma_N^{(2)} \right)^{-1}=- \left(\Gamma_N^{(2)} \right)^{-1} \left( \frac{\partial}{\partial \phi} \Gamma_N^{(2)}\right) \left(\Gamma_N^{(2)} \right)^{-1} $, cf. fig.~\ref{wavefunction}.

Thus the contributions to the $\beta$ function of a particular coupling with $n$ powers of $\phi$ consist of all one-loop diagrams with $n$ external $\phi$'s and therefore up to $n$ vertices. 

The main difference between the matrix model and the usual setting for quantum field theories on a spacetime is the lack of a kinetic operator, and the nonstandard assignment of canonical dimensionality.

As quantum fluctuations generate all operators that are compatible with the symmetries, $\Gamma_N$ will contain further operators beyond those in the bare action. We thus have to consider the Renormalization Group flow in the typically infinite-dimensional theory space of all couplings compatible with the symmetries of the model. While the Wetterich equation is exact, in practice its solution requires a truncation of $\Gamma_N$, which implies an approximative result.
In the present setting, the effective action can be expanded in terms of multiple traces of $\rho = \phi^2$, containing operators such as $\tr \phi^n$ and $\tr \phi^n \tr \phi^m$, which yield a basis for theory space. In our case, the $Z_2$ symmetry restricts the class of allowed operators to those with even powers in the field.
Note that the introduction of the regulator into the generating functional breaks the $U(\Lambda)$ symmetry of the matrix model, as matrix entries below $N$ acquire a mass-type term, while those above $N$ do not. This implies that further terms will be generated that go beyond that theory space, very much in the same way as the flow equation generates, e.g., a gluon mass term in the flow for non-Abelian gauge theories, see, e.g., \cite{Gies:2006wv}. Here, we will only consider the flow of operators that are permitted by the original symmetry \footnote{In principle, the Renormalization Group flow of the hermitian matrix model could be derived after diagonalizing the action. This would however require a technically involved treatment of the Vandermonde determinant, see \cite{Di Francesco:1993nw, Ginsparg:1993is}.}.

To project onto the couplings of the corresponding operators, we will use the $\mathcal{P}^{-1}\mathcal{F}$ expansion of the flow equation: We split $\Gamma_N^{(2)}+R_N = \mathcal{P}_N+\mathcal{F}_N$, where all field-dependent
terms enter the fluctuation matrix $\mathcal{F}_N$. We can then expand the right-hand side of the flow equation as follows:
\begin{eqnarray}
 \partial_t \Gamma_N&=& \frac{1}{2}\textrm{tr} \{
 [\Gamma_N^{(2)}+R_N]^{-1}(\partial_t R_N)\}\label{eq:flowexp}\\
&=& \frac{1}{2} \textrm{tr}\, \tilde{\partial}_t\ln
\mathcal{P}_N
+\frac{1}{2}\sum_{n=1}^{\infty}\frac{(-1)^{n-1}}{n} \textrm{tr}\,
\tilde{\partial}_t(\mathcal{P}_N^{-1}\mathcal{F}_N)^n,
\nonumber
\end{eqnarray}
where the derivative $\tilde{\partial}_t$ in the second line by definition
acts only on the $N$ dependence of the regulator, $\tilde{\partial}_t=
\textrm{tr} \partial_t R_N\frac{\delta}{\delta R_N}$. As each power of $\mathcal{F}_N$ contains a dependence on the matrix $\phi$, this expansion corresponds to an expansion in the number of vertices. To project, e.g., on the running of $g_4$, we need only up to the second order of the $\mathcal{P}^{-1}\mathcal{F}$ expansion, as a tadpole vertex $\sim g_6$ and a two-vertex diagram with vertices $\sim g_4$ are the only one-loop diagrams that can contribute to the running of the $\phi^4$ term. (If our action contained terms $\sim \phi^3$, three-vertex and four-vertex diagrams would also contribute.)

We will now explain how to apply this equation in a concrete case, and show that the perturbative results by \cite{Brezin:1992yc, Ayala:1993fj} can be reproduced.
As a first step, we use that any hermitian matrix $\phi$ can be decomposed into a real symmetric matrix $A$ and a real antisymmetric matrix $B$ by
\be
\phi_{ab} = A_{ab} + i\, B_{ab},
\ee
and decompose all terms ${\rm tr}\phi^n$ accordingly. For the projection onto the invariants ${\rm tr}\phi^i\, \dots {\rm tr} \phi^j$ we can then use diagonal matrices of the form
\be
A_{ij}= a \delta_{ij}.\label{projection}
\ee
Different orders in $\phi$ are then distinguished by powers of $a$. Operators with the same number of fields but different numbers of traces are distinguished by powers of $N$. 
$\Gamma^{(2)}$ then takes the form of a matrix in field space, with entries $\Gamma^{(2)}_{AA}$,  $\Gamma^{(2)}_{AB}$,  $\Gamma^{(2)}_{BA}$ and  $\Gamma^{(2)}_{BB}$, where we can set $B=0$ after taking the derivative.
The inverse propagator matrix then takes a very simple form: Due to the antisymmetry, any term with an odd number of $B$'s vanishes. The two off-diagonal terms  $\Gamma^{(2)}_{AB}$ and  $\Gamma^{(2)}_{AB}$ are also zero, since they contain only terms with odd numbers of $B$s. $\Gamma^{(2)}_{BB}$ is nonzero, but does not contain any powers of $B$. This diagonal structure greatly simplifies the evaluation of the operator trace.

For the derivatives we use that
\bea
\frac{\delta}{\delta A_{ab}}A_{cd} &=& \frac{1}{2}\left(\delta_{ad}\delta_{bc}+ \delta_{ac}\delta_{bd} \right),\\
\frac{\delta}{\delta B_{ab}}B_{cd}&=&-\frac{1}{2}\left(\delta_{ad}\delta_{bc}- \delta_{ac}\delta_{bd} \right).
\eea

The propagators for the two matrix modes are then given by
\bea
P^{-1}_{AA\, abcd} &=& \frac{1}{Z_{\phi}(1+R_N(a,c))} \left(\delta_{ad}\delta_{bc}+ \delta_{ac}\delta_{bd} \right), \\
P^{-1}_{BB\, abcd} &=& -\frac{1}{Z_{\phi}(1+R_N(a,c))} \left(\delta_{ad}\delta_{bc}- \delta_{ac}\delta_{bd} \right), \\
\eea
where $Z_{\phi}$ is a wave function renormalization.

It is important to realize that the derivative of ${\rm tr}  \phi^n$ generates two very different types of terms, only one of which generates nonzero contributions to $\beta_{g_i}$: As an example, consider
\bea
&{}&\frac{\delta}{\delta A_{ab}} \frac{\delta}{\delta A_{cd}} \frac{\bar{g}_4}{4} {\rm Tr} A^4 \nonumber\\
&=& \frac{\bar{g}_4}{2} \Bigl(A_{cn}A_{na} \delta_{bd} + A_{cn}A_{nb}\delta_{ad}+A_{dn}A_{na}\delta_{bc}+A_{dn}A_{nb}\delta_{ac} \nonumber\\
&{}&+ A_{bd} A_{ca}+A_{bc}A_{da} \Bigr),
\eea
where we use a summation convention for indices that occur twice.

Contracting this expression with the propagator and taking the operator trace, there will be a term $\sim \delta_{aa}$ and a term that does not involve a Kronecker-delta anymore:
\bea
&{}&{\rm {tr}} P^{-1}_{AA\, ab cd} \left( \frac{\delta}{\delta A_{ab}} \frac{\delta}{\delta A_{cd}} \frac{\bar{g}_4}{4}{\rm Tr} A^4\right)\nonumber\\
&=& \bar{g}_4 \frac{1}{Z_{\phi}(1+R_N(a,b))} A_{an}A_{na}\delta_{bb} \nonumber\\
&{}&+ \bar{g}_4 \frac{1}{Z_{\phi}(1+R_N(a,a))}A_{an}A_{na}\nonumber\\
&{}&+\frac{\bar{g}_4}{2} \frac{1}{Z_{\phi}(1+R_N(a,b))} A_{ab}A_{ba}\nonumber\\
&{}&+ \frac{\bar{g}_4}{2} \frac{1}{Z_{\phi}(1+R_N(a,b))} A_{aa}A_{bb}
\eea

The terms which are generated correspond to $\tr \left(\phi^2\right)$ and $\left(\tr \left(\phi\right)\right)^2$, since these are the only two terms at $\mathcal{O}(\phi^2)$. The two terms can be clearly distinguished above: The first term will be $\sim \tr\, \phi^2$, as will be the second and third term. The fourth one is clearly $\sim \left(\tr \left(\phi\right)\right)^2$.
Inserting our choice of projection \Eqref{projection} after this identification, we obtain a scaling $\sim N^2$ for the first term and a scaling $\sim N$ for the second and third, and $\sim N^2$ for the fourth one. Going over to the dimensionless coupling $g_4$, each term receives an additional power of $N^{-1}$.
To project onto the dimensionless coupling of $\tr \left(\phi^2\right)$, we require a scaling $\sim N$. Therefore, only the first term will yield a contribution at $\mathcal{O}(N^0)$, whereas the second and third one will be suppressed by $\frac{1}{N}$. To project onto the dimensionful coupling of $\left(\tr \left(\phi\right)\right)^2$, we require a scaling $\sim N$, since the coupling has dimensionality $-1$, and a factor $N^2$ arises from the two traces. Thus the third term gives a $\mathcal{O}(N^0)$ contribution to the flow of this coupling.

We use a regulator that is inspired by Litim's optimized cutoff \cite{Litim:2001up} and takes the form
\be
R_N (a,b)_{abcd} =Z_{\phi} P_{A/B\, abcd} \left( \frac{2N}{a+b} -1\right) \theta \left(1-\frac{a+b}{2N} \right),\label{regulator}
\ee
where $P_{A\,abcd} = \frac{1}{2}\left(\delta_{ad}\delta_{bc}+ \delta_{ac}\delta_{bd} \right)$ for the $A$ mode and $P_{B\,abcd} = -\frac{1}{2}\left(\delta_{ad}\delta_{bc}- \delta_{ac}\delta_{bd} \right)$ for the $B$ mode. The factor of 2 in the argument of the shape function accounts for the fact that both matrix indices $a,b$  will run up to $N$. Here one can observe a major difference to settings with a non-trivial kinetic term, where the corresponding operator would multiply the shape function on the RHS of \Eqref{regulator} to define the regulator.
We then have that
\bea
&{}&\partial_t R_N(a,b)_{abcd}\nonumber\\
& =&  P_{A/B\, abcd} \Bigl[\left(\partial_t Z_{\phi} \left( \frac{2N}{a+b} -1\right)   +Z_{\phi} \frac{2N}{a+b}\right)\theta \left(1-\frac{a+b}{2N} \right)\nonumber\\
&{}&+Z_{\phi} \left( \frac{2N}{a+b} -1\right) \delta \left(1-\frac{a+b}{2N} \right)\frac{a+b}{2N}
\Bigr],
\eea
where the $\delta$ distribution in the second line does not contribute to sums/integrals over $a,b$ due to the factor in front.
For this choice of cutoff we have that
\be
\sum_{a} \sum_{b}\left(P^{-1}_{a\,b}\right)^n \partial_t R_N \sim N^2 \frac{4 \left(2+n - \eta \right)}{2+3n +n^2}\left(1+\mathcal O(1/N)\right).\label{doubletrace}
\ee
Here we have approximated the sum by an integral, which is correct at leading order in $\frac{1}{N}$.
We are now in a position to derive the $\beta$ functions with the Wetterich equation.

\subsection{$\beta$ functions and fixed-point analysis}
\begin{figure}
\includegraphics[width=0.5\linewidth]{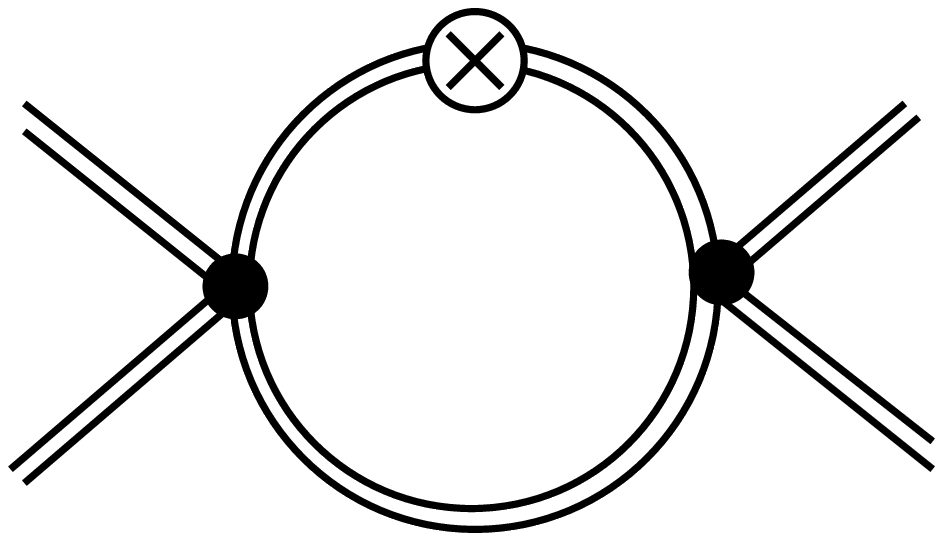} \quad\includegraphics[width=0.25\linewidth]{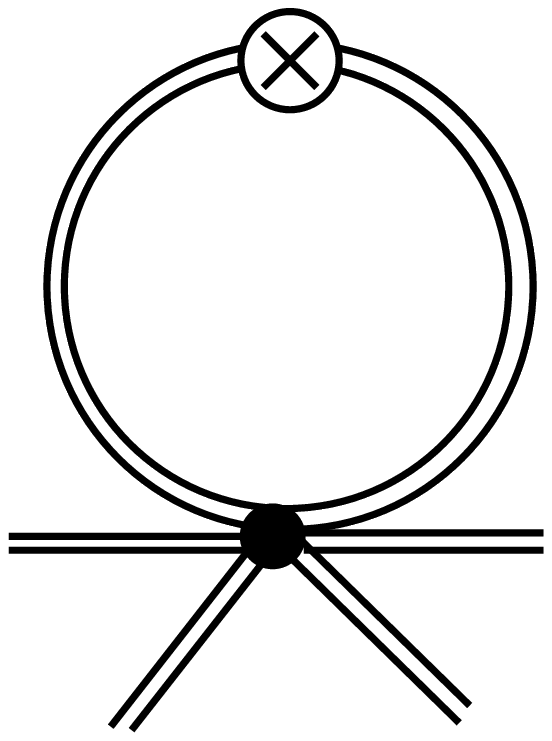}
\caption{\label{betagdiag}A diagram $\sim g_4^2$ and a tadpole diagram $\sim g_6$ contribute to $\beta_{g_4}$.}
\end{figure}

Let us first analyze a simple truncation of the effective action, 
which reads
\be
\Gamma_N = \frac{Z_{\phi}}{2} \tr \left(\phi^2\right) + \frac{\bar{g}_4}{4} \tr \left(\phi^4\right) + \frac{\bar{g}_6}{6} \tr \left(\phi^6\right).
\ee
Using the methods outlined above, we can derive a set of $\beta$ functions, which read
\bea
\eta&:=&- \partial_t \ln Z_{\phi}= 2g_4 \frac{1}{3}\left(4-\eta \right)\label{eta},\\
\beta_g &=& g_4 +2 \eta g_4+ 4 g_4^2 \frac{1}{5} \left(5- \eta \right) -4 g_6 \frac{1}{3}\left(4- \eta \right),\\
\beta_{g_6}&=& 2g_6 +3 \eta g_6 + 12 g_4\, g_6  \frac{1}{5} \left(5- \eta \right) -6 g_4^3 \frac{2}{15} \left(6- \eta \right),\nonumber\\
&{}&
\eea
where the factors of $\eta$ on the right-hand side arise from the use of an RG-adjusted regulator, i.e., $R_N \sim Z_{\phi}$, and accordingly $\partial_t R_N \sim \eta +\dots...$ Diagrammatically, the flow equation for these $\beta$ functions is shown in fig.~\ref{wavefunction} and fig.~\ref{betagdiag}. Similar diagrams with 6 external matrices yield the flow of $g_6$.

The anomalous dimension clearly depends on the coupling in a nonperturbative way, as the solution of \Eqref{eta} reads
\be
\eta= \frac{8 g_4}{3+2 g_4}.\label{etasol}
\ee
We find, apart from the Gau\ss{}ian fixed point, the fixed point corresponding to the double-scaling limit at
\be
g_{4\,\ast} = -0.072, \quad g_{6\, \ast} = -0.004.
\ee
The critical exponents, defined by
\be
\theta_I = - {\rm eig} \left(\frac{\partial \beta_{g_i}}{\partial g_j} \right)\Big|_{g_k = g_{k\, \ast}},
\ee
are given by
\be
\theta_1= 1.046, \quad \theta_2 = -1.080.
\ee
As expected, we obtain one UV-relevant direction, which corresponds to the coupling that has to be tuned in order to reach the double-scaling limit. The exact results are given by
\be
g_c = -\frac{1}{12}, \quad \theta =-0.8.
\ee

To check whether our truncation approaches these exact results, let us consider the smaller truncation with $g_6=0$. Then the fixed point value and critical exponent are given by
\be
g_{4\, \ast}= -0.101, \, \quad \theta = 1.066.
\ee 

We observe that the full result is approached closer in the larger truncation, as expected. We also observe that the nonperturbative information encoded in the structure of $\eta$ through \Eqref{etasol} leads to a smaller value for $g_{4\, \ast}$   in comparison to the results in \cite{Ayala:1993fj}, where $g_{4\, \ast} =-0.103$, $g_{6\, \ast}= -0.005$.

As a next step, we will include further single-trace operators to study the convergence of the critical exponent under extensions of the truncation.

\subsection{Single-trace truncation}\label{sec:singleTraceTruncation}

The single-trace operator truncation, i.e., an ansatz for the effective action of the form
\begin{equation}\label{equ:singleTraceTruncation}
 \Gamma[\phi]=\textrm{Tr}\,V\left(\phi^2\right)=\frac {Z_{\phi}}{ 2} \textrm{Tr}\left(\phi^2\right)+ \sum_{n\ge 2} \frac{\bar g_{2n}}{2n}\,\textrm{Tr}\left(\phi^{2n}\right),
\end{equation}
can be seen as an analogue of the local potential approximation. It shares three features in particular with the local potential approximation: (1) it is a truncation containing an infinite number of couplings  $\bar g_{2n}$, (2) it contains the interesting couplings $g_2=Z_{\phi},\,g_4$ and contains the most relevant coupling at each power of $\phi$, as further traces decrease the canonical dimensionality, in the same way that further derivatives in a standard QFT decrease the canonical dimensionality and (3) one can find a closed expression for the beta functions in the limit $N\to \infty$:
\begin{equation}\label{equ:singleTraceEta}
   \eta = g_4 [\dot R P^2]
\end{equation}
\begin{eqnarray}\label{equ:singleTraceBeta}
&{}&   \beta_{2n} \nonumber\\
&=& \left((1+\eta)n-1\right)g_{2n}\nonumber\\
&+&\!2n\!\!\!\! \!\!\!\sum_{{{i;\, }}n=\sum_i i\,m_i}\!\!\!\!\!\! \left(-1\right)^{\sum_{i} \!\!m_i} [\dot R P^{1+\sum_i \!\!m_i}]
           \left(\begin{array}{c}\sum_i m_i\\m_1,m_2,...\end{array}\right) \prod_{i} g_{2(i+1)}^{m_i},\nonumber
\end{eqnarray}
where the $m_i$ are positive integers {{and $\sum_i m_i$ corresponds to the order of the $\mathcal{P}^{-1}\mathcal{F}$ expansion and denotes the number of vertices in the diagram. By $\dot R P^n$ we denote $\mathcal{P}^{-n} \partial_t R_N$.}}
The bracket denotes the multinomial coefficient $\frac{(\sum_i m_i)!}{m_1!m_2!...}$. {{This factor arises, as in the $\mathcal{P}^{-1}\mathcal{F}$ expansion to order $n$, a contribution proportional to more than one coupling arises several times. For instance, a contribution $\sim g_i g_j$, $i \neq j$ comes with a factor of 2, as either one of the factors $\mathcal{F}$ can be either $\sim g_i$ or $\sim g_j$.}} {{In this expression, the factors $\frac{1}{n}$ from the $\mathcal{P}^{-1}\mathcal{F}$ expansion (cf. \Eqref{eq:flowexp}) cancel against a factor $n$ when the $\tilde{\partial}_t$ derivative acts on $\mathcal{P}^{-n}$.}}
The validity of (\ref{equ:singleTraceBeta}) can be seen in a vertex expansion for the single trace truncation (\ref{equ:singleTraceTruncation}), which starts with the $\mathcal{F}$-term
\begin{equation}\label{equ:SingleTraceFterm}
 F^{abcd}=\sum_{n\ge 1} \bar g_{2n} \left(\delta_{ac}(\phi^2)^{n-1}_{db} + ... + (\phi^2)^{n-1}_{ac}\delta_{db}\right),
\end{equation}
where the ellipsis denotes the summands without any $\delta$. The contribution of any trace that does not contract a complete sequence of $\delta$'s is suppressed by $1/N$ and is thus suppressed in the limit $N\to\infty$. Consequently, there are only two products in $(F^n)^{abcd}$ that are not suppressed in the limit $N\to\infty$, namely the one contracting solely over the initial $\delta$ and the one contracting over the final $\delta$. These terms yield the same field monomial and hence produce a factor of $2$ that cancels the $\frac 1 2$ from the RHS of the flow equation, {{cf. \Eqref{eq:flowexp}}}. Consequently, equation (\ref{equ:singleTraceEta}) for $\eta$ follows from the fact that the only summand in (\ref{equ:SingleTraceFterm}) that contains at most two fields is the term containing a single $g_4$. Moreover, using the reciprocal power series formula 
\bea
&{}&\left(1+\sum_{n\ge1}{{\epsilon}}^n g_n\right)^{-1}\\
&=&\sum_{n=0}^\infty \epsilon^n \sum_{n=\sum_i i m_i} (-1)^{\sum_i m_i}\left(\begin{array}{cc}\sum_i m_i\\m_1,m_2,...\end{array}\right)\prod_i g_i^{m_i},\nonumber
\eea
and dimensionless couplings $g_{2n}=N^{n-1}Z_{\phi}^{-n}\bar{g}_{2n}$, one finds equation (\ref{equ:singleTraceBeta}) directly form the vertex expansion. 

Let us use (\ref{equ:singleTraceEta}) and (\ref{equ:singleTraceBeta}) to investigate the effect of increasing the number of single trace operators in the truncation on the numerical values of the non-Gaussian fixed point. The explicit expressions for the first beta functions are
\begin{eqnarray}
 \beta_4&=&(1+2\eta)g_4+4\left(-[\dot R P^2] g_6+[\dot R P^3] g_4^2\right),\\
 \beta_6&=&(2+3\eta)g_6+6\left(-[\dot R P^2] g_8+2[\dot R P^3] g_4g_6-[\dot R P^4] g_4^3\right),\nonumber\\
&{}&
\eea
\bea
 \beta_8&=&(3+4\eta)g_8+8\left(-[\dot R P^2] g_{10}+[\dot R P^3] (g_6^2+2g_8g_4)\right.\\
               &{}         &\left.-3[\dot R P^4] g_4^2g_6+[\dot RP^5]g_4^4\right),\nonumber\\
 \beta_{10}&=&(4+5\eta)g_{10}\nonumber\\
&{}&+10\left(-[\dot R P^2] g_{12}+2[\dot R P^3] (g_4g_{10}+g_6g_8)\right.\\
                      &{}  &\left.-3[\dot R P^4] (g_4^2g_8+g_4g_6^2)+4[\dot RP^5]g_4^3g_6-[\dot RP^6]g_4^5\right),\nonumber\\
 \beta_{12}&=&(5+6\eta)g_{12}+12\left(-[\dot R P^2] g_{14}\right.\\
&{}& \left.+[\dot R P^3] (2g_4g_{12}+2g_6g_{10}+g_8^2)\right.\nonumber\\
&{}&\left.                    -[\dot R P^4] (g_6^3+3g_4^2g_{10}+6g_4g_6g_8)\right.\nonumber\\
&{}&\left.+[\dot RP^5](6g_4^2g_6^2+4g_4^3g_8)\right.\nonumber\\
                       &{} &\left.-5[\dot RP^6]g_4^4g_6+[\dot R P^7]g_4^6\right),\nonumber\\
 \beta_{14}&=&(6+7\eta)g_{14}+14\left(-[\dot R P^2] g_{16}\right.\\
&{}&\left.+2[\dot R P^3] (g_4g_{14}+g_6g_{12}+g_8g_{10})\right.\nonumber\\
                   &{}     &\left.-3[\dot R P^4] (g_4^2g_{12}+2g_4g_6g_{10}+g_4g_8^2+g_6^2g_8)\right.\nonumber\\
&{}&\left.+4[\dot RP^5](g_4^3g_{10}+3g_4^2g_6g_8+g_4g_6^3)\right.\nonumber\\
                       &{} &\left.-5[\dot RP^6](g_4^4g_8+2g_4^3g_6^2)+6[\dot R P^7]g_4^5g_6-[\dot R P^8]g_4^7\right).\nonumber
\end{eqnarray}
The factors $\dot R P^n$ are dependent on the choice of regulator and defined by \Eqref{doubletrace} for our choice of shape function.
We find a non-interacting Gau\ss{}ian fixed point with canonical critical exponents and several interacting fixed points, of which the only one with only one relevant direction is the one of interest for the double-scaling limit.
The numerical values $g_{2n}^*$ for the couplings at that particular non-Gaussian fixed point and the critical exponents $\theta_i$,  are given in tab.~\ref{fptab}.
\begin{widetext}
\begin{center}
\begin{table}[!here]
\caption{\label{fptab}Critical exponents $\theta_i$ and fixed point values at $\mathcal{O}(n)$ in the $\mathcal{P}^{-1}\mathcal{F}$ expansion.}
\begin{tabular}{c|cccccc|cccccc||}
  $n_{max}$ & $\theta_1$ & $\theta_2$& $\theta_3$& $\theta_4$& $\theta_5$& $\theta_6$& $g_4^*$ & $g_6^*\cdot 10^3$ & $g_8^* \cdot 10^4$ & $g_{10}^*\cdot 10^5$ & $g_{12}^*\cdot10^5$ & $g_{14}^*\cdot 10^6$\\ \hline
  2 & 1.066 & & & & & & -0.1005&  &  & & &\\
  3 & 1.046 &-1.080& & & & & -0.0722&  -3.8&  & & & \\
  4 & 1.036 &-1.053& -2.14& & & & -0.0563&  -4.6&  -3.6& & & \\
  5 & 1.029 & -1.037&-2.115&-3.171& & &-0.0461&  -4.7&  -5.4& -5 & &\\
  6 & 1.025 & -1.027&-2.10&-3.137&-4.197& & -0.0390&  -4.5&  -6.2& -8 & -1 &\\
  7 & 1.022 & -1.020&-2.093&-3.110&-4.172&-5.213&-0.0338&  -4.2&  -6.5&-10 & -1 & -1.5\\
  \hline
\end{tabular}
\end{table}
\end{center}
\end{widetext}
It turns out that in each order in the expansion there is only one positive critical exponent, $\theta_1$, which means in particular that increasing the truncation does not introduce new relevant directions.
We also observe that the irrelevant critical exponents approach the values $\theta_i =d_i+1$, where $d_i$ is the canonical dimensionality of the couplings in the truncation. At the same time, the eigenvectors of the stability matrix do not directly correspond to the couplings $g_i$, but to superpositions of them. To increasing order in the truncation, we observe an apparent convergence of the eigenvalues:
We observe that the positive critical exponent approaches smaller values. As the difference in the values between subsequent truncation orders becomes smaller in every step, an extrapolation of $\theta$ for truncation order $n \rightarrow \infty$ yields $\theta>1$. We conclude that operators are still missing from this truncation. Very much in analogy to the critical exponents for the Wilson-Fisher fixed point, further operators at each order in the fields are missing: For the Wilson-Fisher fixed point, next-to-next to leading order in the derivative expansion is necessary for accurate critical exponents \cite{Canet:2003qd,Litim:2010tt}. In our case, the missing operators at a given power of the fields are multi-trace operators. We will demonstrate how to include these in the following. Let us briefly explain why we take into account operators that are power-counting irrelevant, cf. \Eqref{doubletracescaling}, and that we expect to remain irrelevant, as the double-scaling limit corresponds to only one 
relevant direction. At an interacting fixed point, both irrelevant and relevant couplings take non-zero values. Accordingly, nonvanishing contributions from the fluctuation of irrelevant operators couple into the flow of relevant couplings. Thus the fixed-point values and critical exponents of a relevant direction receive nonvanishing contributions from irrelevant couplings. Therefore the (ir) relevance of a coupling is {\emph{not}} a criterion to distinguish whether this coupling yields large correction to $\beta$ functions at an interacting fixed point.

\subsection{Double-trace operators and the RG flow}

We observe that operators of the form $\tr \phi^2 \tr \phi^i$ couple directly into the flow of $g_i$ by a tadpole diagram.
This can be seen as follows: In order to contribute to $\beta_{g_i}$, the corresponding diagram must come with an overall factor of $N^{\frac{-i+2}{2}}$. For the above operator, the tadpole diagram contributes with a factor of $N^2$ from the evaluation of the trace. This can be seen directly, as the second variation of $\tr \phi^2 \tr \phi^i$ contains terms of the form $\tr \phi^i P_{A/B\, abcd}$. After contracting this with the propagator, $\tr \phi^i$ can be pulled out of the trace, and a trace over $P_{A/B\, abcd}$ remains, which is $\sim N^2$, cf. \Eqref{doubletrace}. Going over to the dimensionless coupling $g_{2,i}$ yields another factor of $N^{\frac{-i-2}{2}}$, thus contributing to $\partial_t g_i$.\\
We concluce that operators with $n-1$ traces are generated from $n$- trace operators of the form $\tr \phi^2 \tr \phi^i \dots \tr \phi^j$. The contribution of further $n+1$ trace operators not containing a term $\tr \phi^2$ is suppressed by $\frac{1}{N}$.

We thus extend our truncation by the two two-trace operators that couple directly into $\beta_{g_4}$ and $\eta$ and consider:
\bea
\Gamma_N &=& \frac{Z_{\phi}}{2} \tr \left(\phi^2\right) + \frac{\bar{g_4}}{4} \tr \left(\phi^4\right) + \frac{\bar{g}_6}{6} \tr\left( \phi^6\right)\nonumber\\
&{}&+ \frac{\bar{g}_{2,2}}{4} \left(\tr \left(\phi^2\right) \right)^2 + \frac{\bar{g}_{2,4}}{2} \tr \left(\phi^2\right) \tr \left(\phi^4\right).
\eea

For this truncation we obtain the following beta functions:
\bea
\eta&=& (2g + g_{2,2}) [\dot R P^2]\\
\beta_g &=& g + 2 \eta g +  4 g^2 [\dot R P^3] \nonumber\\
&{}&- 4 g_6 [\dot R P^2] -2 g_{2,4}[\dot R P^2]\\
\beta_{g_{2,2}}&=& 2 g_{2,2}+ 2 \eta g_{2,2} + 8 g g_{2,2} [\dot R P^3] \nonumber\\
&{}&+2 g_{2,2}^2 [\dot R P^3] -8 g_{2,4} [\dot R P^2]\nonumber\\
&{}&-2 g_6 [\dot R P^2]+6 g^2 [\dot R P^3]\\
\beta_{g_6}&=&2g_6 +3 \eta g_6 + 12 g g_6 [\dot R P^3]\nonumber\\
&{}& - 6 g^3 [\dot R P^4]
\eea
\bea
\beta_{g_{2,4}}&=& 3 g_{2,4}+ 3 \eta g_{2,4} +4 g_{2,2}g_6 [\dot R P^3] \\
&{}&+12 g g_{2,4} [\dot R P^3] \partial_t R\nonumber\\
&{}&+ 2 g_{2,2} g_{2,4} [\dot R P^3] \partial_t R-12 g^3 [\dot R P^4] \partial_t R\nonumber\\
&{}&+12 g g_6 [\dot R P^3] -6 g^2 g_{2,2}[\dot R P^4].\nonumber
\eea

Using the regulator \Eqref{regulator} and employing \Eqref{doubletrace} we obtain several interacting fixed points and the non-interacting Gau\ss{}ian fixed  point. For the fixed point with only one relevant direction we get the following values:

\bea
g_{\ast}&=& -0.056, \quad g_{6\, \ast} = -0.0015, \quad g_{2,2\,\ast}= -0.058, \nonumber\\
 g_{2,4\, \ast}&=& -0.0027,\\ \nonumber
\theta_1&=& 1.21, \quad \theta_2=-0.69, \quad \theta_3=-1.01, \quad \theta_4= -1.88.
\eea
These values should be compared to the perturbative result, where $g_{\ast} =-0.101$  and $\theta_1=1.22$ \cite{Ayala:1993fj}. The nonperturbative effects constitute a slight improvement over the perturbative ones. 

We expect an improvement of the critical exponents at higher orders of the truncation, when further double-trace operators and also triple-trace operators, which couple into the flow of the double-trace ones, are taken into account. As the goal of the present work is to establish the functional RG as a novel tool, we do not embark on an extended study here to attain quantitative precision. The satisfactory agreement of the results obtained with our method with those in the perturbative study show that our method works well.

\subsection{Equivalence of the even hermitian matrix model with a real matrix model}

A first step from the even hermitian matrix model towards tensor models is to consider a real matrix model that is invariant under a pair of orthogonal transformations
\begin{equation}
 \phi \to O_1^T \phi\, O_2,
\end{equation}
where $O_1$ and $O_2$ are arbitrary orthogonal matrices. The invariants of this model are sums and products of traces of arbitrary powers of $\rho=\phi^T\phi$. This gives the following theory space
\begin{equation}
 \Gamma[\phi]=\frac {Z_{\phi}}{ 2} \textrm{Tr}\left(\rho\right)+\sum_{i\ge 1}\sum_{n_1,...,n_i\ge 0}\bar g_{n_1,...,n_i}\textrm{Tr}\left(\rho^{n_1}\right)...\textrm{Tr}\left(\rho^{n_i}\right).
\end{equation}
Our explicit calculations of the truncations presented in this paper for the hermitian model show that the large N limit of the beta functions of this model coincides with the large N limit of the beta functions of the $Z_2$ symmetric hermitian matrix model if we identify operators of the two models using the correspondence
\begin{equation}
 \phi^T\phi \textrm{ (real bi-orthogonal) } \leftrightarrow \phi^2 \textrm{ ($Z_2$ symmetric hermitian).}
\end{equation}

\section{Colored matrix model}\label{cmm}
We will now consider a colored matrix model, inspired by the colored tensor models of \cite{Gurau:2009tw} to demonstrate how the Wetterich equation works in this case. 

Consider a real $N\times N$ matrix model with three colors, i.e., the fundamental degrees of freedom are $\phi_i$ with $i=1,2,3$, and tri-orthogonal invariance, i.e., invariance under
\begin{equation}
 \phi_i \to O^T_i\, \phi_i \,O_{(i+1)\textrm{mod}\,3},
\end{equation}
where $O_i\in O(N)$. This model has bi-linear invariants of the form
\begin{equation}
 x^{(i)}=\textrm{Tr}\left(\phi_i^T\phi_i\right),
\end{equation}
a tri-linear invariant of the form
\begin{equation}
 s=\textrm{Tr}\left(\phi_1\phi_2\phi_3\right),
\end{equation}
and higher invariants. These higher invariants are traces of strings of $\phi_{i}...$'s satisfying the rules
\begin{enumerate}
 \item to the right of $\phi_i$ is either $\phi_{(i+1){\rm mod} 3}$ or $\phi_i^T$
 \item the beginning of a string ending with $\phi_i$ is either $\phi_{(i+1)){\rm mod} 3}$ or $\phi_i^T$.
\end{enumerate}
Cyclicity and invariance of the trace under transposition imply that two invariants which are only cyclic permutations or transposition of each other represent the same invariant. To have a representation of the invariants we define the map $\mu(\phi^T_i)=i-1$ and $\mu(\phi_i)=i+2$, so we associate to each string of $\phi...\phi$'s an integer with base-6 representation $\mu(\phi)...\mu(\phi)$. Then for each invariant we take the representative string with the smallest associated integer. This leads to the following ansatz for the vertex expansion:
\begin{equation}
 \Gamma_k[\phi]=\sum_{n=1}^{\infty} \sum_{i_1,...,i_n\in\mathcal I}\sum_{j_1,...,j_n=1}^\infty g^{j_1...j_n}_{i_1...i_n}\textrm{Tr}\left(S_{i_1}\right)^{j_1}...\textrm{Tr}\left(S_{i_1}\right)^{j_1},
\end{equation}
where $\mathcal I$ denotes the set of minima over permutations of base 6-digits of an integer and $S_i$ the string associated to the integer $i$.

Our ansatz for the canonical dimensionality of the couplings again follows from \Eqref{canondim}.
Let us consider the following truncation
\be
\Gamma_N = \frac{Z_{\phi}}{2}\sum_{i=1}^3  \tr \left(\phi_i^T\phi_i\right)+ g\, \tr (\phi_1 \phi_2 \phi_3 +\phi_1^T\phi_3^T\phi_2^T).
\ee
The inverse propagator is given by
\be
 P^{ij}_{abcd}= Z_{\phi} \delta_{ac} \delta_{bd}\delta_{ij},
\ee
i.e., it is diagonal in the colors. The fluctuation matrix $\mathcal{F}$ takes a particularly simple form, with only off-diagonal entries and $\mathcal{F}_{12} \sim \phi_3$ etc. The anomalous dimension is then calculated from a two-vertex diagram $\sim g^2$ and reads
\be
\eta= 8 g^2 (1- \eta/3).
\ee
Interestingly, we observe that there is no contribution $\sim g^3$ to $\beta_g$, as the corresponding three-vertex diagram is suppressed by $\frac{1}{N}$. Accordingly, the $\beta$ function at this order reads
\be
\beta_g = \frac{1}{2}g + 3 \eta g = g\left(\frac{1}{2} +24^2 \frac{g^2}{3+8 g^2} \right).
\ee
Within this perturbative expansion, the model only admits a UV-repulsive Gau\ss{}ian fixed point. 
The critical exponent at this fixed point is $\theta= -1/2$, thus the fixed point is UV-repulsive. Such a behavior is well-known in Quantum Electrodynamics, which has a similar UV-repulsive Gau\ss{}ian fixed point, and exhibits a Landau pole and a triviality problem.
At higher order in the couplings, the operator $\tr \left(\phi_1 \phi_2 \phi_3 \phi_1 \phi_1^T\right)$ can generate a further contribution. This is expected to shift the Gau\ss{}ian fixed point to become an interacting fixed point and imply a correction to the critical exponent arising from quantum fluctuations. 

A similar effect occurs in colored tensor models in higher dimensions, since for instance the bare action truncation $\frac {Z_{\phi}}{ 2} \bar \phi^{(i)} \phi^{(i)} + g \left(\phi^{(1)}\phi^{(2)}\phi^{(3)}\phi^{(4)}+ h.c.\right)$ cannot produce the interaction term at one loop. The renormaliztion of $g$ has to be mediated through effective operators.

\section{Recipe}\label{recipe}

The purpose of this paper is in part to establish the functional Renormalization Group equation as a useful tool in the study of models that are technically similar to modern tensor models. It is useful, for this wider purpose, to extract a recipe for the treatment of matrix models, that can serve as the starting point for generalizations. In particular, the generalization to tensor models with a "trivial" kinetic term, similar to the hermitian matrix model, is straightfoward, as only the number of indices changes.

\subsection{Theory Space and Truncations}

The first step in the setup of a flow equation is the specification of a theory space. The theory space is the space of all operators compatible with the symmetries. As long as the choice of regulator does not break the symmetries, no operators beyond this theory space can be generated by the flow equation.
To define the theory space one often starts from a bare, i.e., microscopic, action $S[\phi]$ that one would like to have in the theory space and specifies the field content of this bare action and its linearly realized symmetries that can be implemented as symmetries of the regulator\footnote{Symmetries that are not linearly realized or can not be implemented in the regulator will not become symmetries of the effective average action, but rather lead to Ward identities that are very hard to solve in practice.}.

In the present paper, we are concerned with matrix models, so the field content is a multiplet of matrices $\phi^I_{ab}$ with possible restrictions on the matrices, e.g., real, real symmetric or hermitian. The linearly realized symmetries can be, e.g., $Z_2$-invariance under $\phi^I_{ab} \to -\phi^I_{ab}$ or invariance under ``internal'' 
transformations $\phi^I_{ab} \to M^I_J \phi^J_{ab}$ or ``local'' symmetries $\phi^I_{ab} \to T^I_{ac}\phi^I_{cd}R^I_{db}$. The ansatz for the effective average action is then a functional $\Gamma_k[\phi^I]$ that is invariant under these symmetries {{and can be written as a sum of basis operators in theory space, with scale-dependent running couplings as prefactors.}}

 For instance the $Z_2$-symmetric hermitian matrix model considered in this paper leads to the following ansatz for the effective average action:
\begin{equation}\label{equ:TheorySpaceEvenHermitian}
 \Gamma_k[\phi]=F_k\left(\textrm{Tr}\left(\phi^2\right),\textrm{Tr}\left(\phi^4\right),...\right)
\end{equation}
where the field content is given by a single hermitian matrix $\phi$.

The theory space for tensor models is constructed in complete analogy: One specifies a set of tensors $\phi^I_{a_1...a_n}$, which do not have to all be of the same rank or type and a set of linearly realized symmetries, which leads to an ansatz for the effective average action analogous to equation (\ref{equ:TheorySpaceEvenHermitian}).

It is generally unfeasible to treat the entire theory space {{as it is infinite dimensional}}. Thus one is forced to make more restrictive ans\"atze for the effective average action, i.e., {{specify}} truncations, which are generally not preserved by the flow equation. {{Hence the quantum fluctuations of operators within the truncation generate nonvanishing flows for operators outside the truncation and conversely not all contributions to the running of operators within the truncation are necessarily included. To devise a good truncation, one should include the operators that carry the main contribution to the running of the couplings of interest.}} Note that even infinite-dimensional truncations of the Wetterich equation can be treated, see, e.g., \cite{Gies:2002af, Benedetti:2012dx,Dietz:2012ic} for examples in gauge theories and gravity.
Pure matrix models offer a good way to study the accuracy of a truncation, because, in a vertex expansion, there is only a finite number of operators that can contribute to the running of any given operator in the truncation. This means that one can enlarge any finite truncation to a larger finite truncation that contains all operators that can contribute to the running of the original truncation and study the effect of the new operators. For example, in the case of the hermitian matrix model one can start with the bare action as the first truncation of the effective action
\begin{equation}\label{equ:bareTruncation}
 \Gamma^{(0)}_k[\phi]=\frac {Z_{\phi}} {2} \tr\left(\phi^2\right)+\frac{g_4}{4}\tr\left(\phi^4\right).
\end{equation}
We now use that in the vertex expansion the running of an operator with $n$ $\phi$'s {{and $m$ traces}} can only be influenced by operators with up to $n+2$ $\phi$'s. {{Furthermore}}  the only operator with more than $m$ traces that can influence the running of an operator $O[\phi]$ with $m$ traces in the large $N$-limit is $O[\phi]\,\textrm{Tr}\left(\phi^2\right)$. To include all these operators, we enlarge the truncation (\ref{equ:bareTruncation}) to
\bea
 \Gamma^{(1)}_k[\phi]&=&\Gamma^{(0)}[\phi]+\frac{g_6}{6}\tr\left(\phi^4\right)+\frac{g_{2,2}}{4}\left(\tr\left(\phi^2\right)\right)^2\nonumber\\
&{}&+\frac{g_{2,4}}{2}\tr\left(\phi^2\right)\tr\left(\phi^4\right).
\eea
The feature that only a finite number of operators can contribute to the running of a given operator is a consequence of the vertex expansion and the fact that a pure matrix model does, by definition, not contain any analogue of ``derivative'' operators, which would be given by constant matrices whose powers could appear anywhere within a trace. The feature thus generalizes to pure tensor models that admit an analogous vertex expansion. {{In the case where derivatives exist, an expansion in the number of derivatives often works very well and can yield quantitatively precise results from the flow equation already at low order in the derivative expansion, see, e.g., for the case of the Wilson-Fisher fixed point in 3-dimensional scalar models \cite{Canet:2003qd,Litim:2010tt}.}}

\subsection{Canonical Dimension}

Pure matrix models lack the usual dimensions of momenta, since these models do, by definition, not posses any analogue of derivative operators. We are however in particular interested in scaling properties of a matrix model with $N$-dependent bare action $S$ when the matrix size $N$ is increased. If one now chooses an appropriate regulator and inserts the bare action into the flow equation then one will in general generate an infinite number of new operators on the RHS of the flow equation. The coefficient of an operator $O[\phi]$ on the RHS of the flow equation will appear with a leading power $N^a$. Thus, if we want to take a sensible $N\to\infty$ limit, we have to scale the bare coupling that corresponds to $O[\phi]$ at least with $N^a$. One then has to iterate the procedure until one has found {{a}} consistent scaling for all bare couplings . This iteration of course requires the use of the scaling bound found from the 
RHS of the previous iteration step on the LHS of the flow equation. The iteration stops when the scaling of the entire theory space on the LHS of the flow equation is consistent with the scaling on the RHS. For instance, in {{the case of }} the  {{$Z_2$ symmetric}} hermitian matrix model, we found that the scaling
\begin{equation}
  \bar g_{2n_1,...,2n_i}N^{2-i+\sum_{k=1}^i n_k}= g_{2n_1,...2n_i},
\end{equation}
allows one to take the limit $N\to\infty$ and yields the scalings for $g_2$ and $g_4$ that were desired from the point of view of random discretizations.

The procedure we outlined here can in principle be applied to pure tensor models as well. The iteration procedure is however technically more complicated, because in colored models many operators are not generated from a bare action at one loop, and thus not directly generated on the RHS of a bare action truncation. 

\subsection{Choice of Regulator}

There is no natural discrimination between IR and UV degrees of freedom in a pure matrix model. In fact, e.g., the unitary symmetry of the hermitian matrix model implies that all degrees of freedom should be treated on the same footing, very similar to gauge theories, where the gauge transformation mixes UV and IR degrees of freedom. This is a linear symmetry, generated infinitesimally by
\begin{equation}
 G_\epsilon\phi_{ab} \to \phi + \epsilon [H,\phi]_{ab},
\end{equation}
where $H$ is a hermitian generator. This linear symmetry can be implemented in the flow equation if we use a scale-independent mass term as IR-regulator, which suppresses all degrees of freedom with the same strength. This is problematic in an infinite matrix model, where we should rather implement a cut-off that depends on matrix size, so we can use the flow equation to study the behavior of the model when the matrix size is increased. This can be done using \Eqref{DeltaS}. 

The dependence of $R_{a,b}$ on the matrix indices $a,b$ breaks the unitary symmetry of the system and modifies the associated Ward-identity \cite{Litim:1998wk,Litim:1998nf} to a complicated nonlinear expression
\begin{equation}
 G_\epsilon \Gamma = \frac 1 2 \textrm{tr}\left(\frac{G_\epsilon R}{\Gamma^{(2)}+R}\right),
\end{equation}
which tells us that we should not construct our theory space using field monomials that are invariant under unitary transformations. Instead we should work with truncations that additionally involve index-dependent operators and (at least approximately) solve the modified Ward-identities. This is however practically unfeasible. Moreover, our direct calculations in the hermitian matrix model, in which we ignored this issue, show good agreement with exact results. So it seems that the modification to the Ward-identity can be neglected for first investigations. 

 The requirements \Eqref{IRsup}, \Eqref{IRlim} and \Eqref{UVlim} that a valid choice of regulator has to satisfy can be generalized to tensor models.
Using an IR suppression term
\be
 \Delta_N S = \frac 1 2 \sum_{a_i, b_i} \phi_{a_1\dots a_k}R_N(a_1,...,a_k)_{a_1\dots a_k\, b_1 \dots b_k} \phi_{b_1\dots b_k},
\ee
with analogous requirements for $\Delta_N S$.

\subsection{Vertex Expansion}
An important tool in the calculations in this paper is the vertex expansion, which is a  standard technique in one-loop calculations. In its simplest form, it can be viewed as a consequence of the fact that $\Gamma^{(2)}[\phi]+R$ can be written as a sum of a field-independent term $\mathcal{P}:=\Gamma^{(2)}[\phi\equiv 0]+R$  and a field-dependent term $\mathcal{F}[\phi]=\Gamma^{(2)}[\phi]-\Gamma^{(2)}[\phi\equiv 0]$. The expansion of the RHS of the flow equation around $\phi\equiv 0$ is thus given by \Eqref{eq:flowexp}. 

This is particularly useful for the derivation of the running of operators that are polynomials in the fields, if, as we assume throughout, $\Gamma^{(2)}[\phi]$ admits a Taylor expansion around $\phi\equiv 0$. The summands in the vertex expansion are 
\begin{equation}
 (-1)^{n+1}\dot R_{ab,cd} \mathcal{P}^{-1}_{cd,r_1s_1} \mathcal{F}[\phi]_{r_1s_2,r_2s_2}...\mathcal{P}^{-1}_{cd,r_ns_n} \mathcal{F}[\phi]_{r_ns_n,tu}\mathcal{P}^{-1}_{tu,ab}.
\end{equation}
In an approximation that ignores index-dependent operators, one can ignore the commutators $[\mathcal{P},\mathcal{F}]$ and the summands in the vertex expansion simplify to
\begin{equation}
 (-1)^{n+1}(\dot R\left( \mathcal{P}^{-1}\right)^{n+1})^{ab,cd}(\mathcal{F}^n[\phi])^{cd,ab}.
\end{equation}
If we now insert field configurations $\phi_{ab}=\phi \delta_{ab}$, then the effect of $(\dot R \left( \mathcal{P}^{-1}\right)^{n+1})^{ab,cd}$ is to cut-off the trace $(\mathcal{F}^n[\phi])^{ab,ab}$ and to multiply it with numerical factor $[\dot R P^{n+1}]:=(\dot R \left( \mathcal{P}^{-1}\right)^{n+1})^{ab,ab}$. Denoting the cut-off trace by $[\mathcal{F}^n[\phi]]_N$, allows us to denote the summands in the vertex expansion in a very simple manner
\begin{equation}
 (-1)^{n+1}[\dot R P^{n+1}][\mathcal{F}^n[\phi]]_N.\label{equ:VertexApprox}
\end{equation}
Practically, one can directly calculate the sum $[\dot R P^{n+1}]$ (at least in a $1/N$ expansion). The derivation of $[F^n[\phi]]_N$ on the other hand can be performed directly.

One can of course enlarge the theory space and include index-dependent operators, generated by $\dot R$ and $\mathcal{P}$. In this case, one expresses $\dot R^{ab,cd} P^{cd,r_1s_1} F[\phi]^{r_1s_2,r_2s_2}...P^{cd,r_ns_n} F[\phi]^{r_ns_n,tu}P^{tu,ab}$ directly in terms of matrix traces with insertions of linear operators that produce the index dependence. This omits the approximation described so far. 

The vertex expansion can be straightforwardly generalized to tensor models, provided one has an invertible quadratic term. The only change that occurs in this case is that the number of indices increases, in particular one proceeds as above but using $\dot R^{a_1...a_n,b_1...b_n}$, $P^{a_1...a_n,b_1...b_n}$ and $F[\phi]^{a_1...a_n,b_1...b_n}$, whose products produce contraction patterns among the fields $\phi^{a_1...a_n}$ that are more general than matrix traces.

\subsection{Extraction of Beta Functions}
Using the vertex expansion as we described it in the approximation leading to equation (\ref{equ:VertexApprox}), it is straightforward to extract beta functions by simply regarding the cut-off traces as matrix traces. The left- and right hand side of the flow equation are then both expanded in terms of products of traces of field products of the form $\textrm{Tr}\left(\phi^{n_1}\right)...\textrm{Tr}\left(\phi^{n_k}\right)$, so one can read-off the beta functions for the bare couplings and one obtains the beta functions in the standard way normalizing the bare couplings with the cut-off times canonical dimension and the appropriate power of the wave function normalization, giving beta functions of the form of \Eqref{equ:BetaType}. The application to tensor models is straightforward once the canonical dimension and wave function normalization are determined in the model.

\section{Conclusions}\label{conclusion}
We have established the nonperturbative functional Renormalization Group as a novel method to study the continuum limit of matrix models for quantum gravity and extracted a recipe for the generalization to tensor models. Building on the insight in \cite{Brezin:1992yc}, where the double-scaling limit implies a Renormalization Group flow of the couplings with the matrix size $N$, we implement a Wilsonian integration over matrix entries. We set up a functional Renormalization Group equation, generalizing the Wetterich equation to this new setting. As these models do not have a Laplacian, the usual integration over fluctuations fields sorted by the eigenvalue of the kinetic term is replaced by an integration over matrix size $N$, with $N$ playing the role of the momentum cutoff $k$ in standard QFTs. We establish a method to determine the canonical dimensionality of couplings, i.e., their canonical scaling with $N$.

As in the standard Wilsonian Renormalization Group flow, further matrix operators are generated in the effective action, beyond those present in the microscopic action. We establish an expansion in the power of the matrix and the number of traces -- analogous to the derivative expansion in a standard quantum field theory -- as a useful expansion for the Renormalization Group flow.

We show the validity of our method by explicitly reproducing the leading-order results in \cite{Brezin:1992yc}, and then extend the truncation of the operator space to include double-trace operators. Here we show how the RG flow of $n+1$ trace operators is connected to that of $n$ trace operators. We include up to two-trace operators of order $\phi^6$ in our truncation, and evaluate the nonperturbative beta functions for the corresponding five running couplings. We obtain results which compare well with the explicit integration over matrix entries in \cite{Brezin:1992yc, Ayala:1993fj}.

As a next step, we discuss how to extend this method to tensor models. There, fixed points of the Renormalization Group flow again correspond to points at which the continuum limit can be taken. They can also be interpreted in a more direct physical sense as second-order phase transitions from  a pregeometric to a geometric phase. In both cases, the universality class shows how physical quantities will scale in the vicinity of the fixed point. In particular, it is interesting to know how many relevant couplings exist, as these correspond to couplings that need to be tuned in order to reach the continuum limit.

We then study a colored matrix model as a toy model for colored tensor models, and show how to obtain the RG flow. We observe that within a truncation containing only one interaction term, the only contribution to the beta function of the corresponding coupling arises from canonical scaling and from a nontrivial anomalous dimension. At that order, the particular model shows only a UV-repulsive Gau\ss{}ian fixed point.

Our method works in the perturbative regime, to show the property in asymptotic freedom as discovered in several tensor models \cite{BenGeloun:2012pu,BenGeloun:2012yk,Geloun:2012qn}, but also goes beyond and could show whether some of these models are actually asymptotically safe \cite{Weinberg:1980gg} and and whether these models possess an interacting fixed point. This nonperturbative notion of renormalizability is a direct extension of asymptotic freedom, and has been studied in the context of the Standard Model \cite{Gies:2009hq, Gies:2013pma} and particularly a local continuum quantum field theory setting for quantum gravity \cite{Reuter:1996cp,Reuter:2012id}. At the corresponding interacting fixed point, the continuum limit of tensor models could potentially yield a phase with an extended semiclassical geometry.

We provide a general recipe how to apply the functional Renormalization Group to a matrix or tensor model, which can provide a basis for future research on the continuum limit in these models.
\newline\\

{\emph{Acknowledgements}}:
This research was supported in part by Perimeter Institute for Theoretical Physics and in part by NSERC. Research at Perimeter Institute is supported by the Government of Canada through Industry Canada and by the Province of Ontario through the Ministry of Research and Innovation. T.~K. acknowledges the hospitality of Perimeter Institute where this work was completed. A.~E. acknowledges financial support by the DFG-Research Training Group ”Quantum- and Gravitational Fields” (GRK 1523/1) during the initial stages of this work.


\begin{thebibliography}{99}

%\cite{Ambjorn:1991pq}
\bibitem{Ambjorn:1991pq} 
  J.~Ambjorn and J.~Jurkiewicz,
  %``Four-dimensional simplicial quantum gravity,''
  Phys.\ Lett.\ B {\bf 278}, 42 (1992).
  %%CITATION = PHLTA,B278,42;%%

%\cite{Ambjorn:2012jv}
\bibitem{Ambjorn:2012jv} 
  J.~Ambjorn, A.~Goerlich, J.~Jurkiewicz and R.~Loll,
  %``Nonperturbative Quantum Gravity,''
  Phys.\ Rept.\  {\bf 519}, 127 (2012)
  [arXiv:1203.3591 [hep-th]].
  %%CITATION = ARXIV:1203.3591;%%

%\cite{Ambjorn:2011cg}
\bibitem{Ambjorn:2011cg} 
  J.~Ambjorn, S.~Jordan, J.~Jurkiewicz and R.~Loll,
  %``A Second-order phase transition in CDT,''
  Phys.\ Rev.\ Lett.\  {\bf 107}, 211303 (2011)
  [arXiv:1108.3932 [hep-th]].
  %%CITATION = ARXIV:1108.3932;%%

%\cite{Rivasseau:2012yp}
\bibitem{Rivasseau:2012yp} 
  V.~Rivasseau,
  %``The Tensor Track: an Update,''
  arXiv:1209.5284 [hep-th].
  %%CITATION = ARXIV:1209.5284;%%

%\cite{Rivasseau:2011hm}
\bibitem{Rivasseau:2011hm} 
  V.~Rivasseau,
  %``Quantum Gravity and Renormalization: The Tensor Track,''
  AIP Conf.\ Proc.\  {\bf 1444}, 18 (2011)
  [arXiv:1112.5104 [hep-th]].
  %%CITATION = ARXIV:1112.5104;%%


%
\bibitem{Boulatov:1992vp} 
  D.~V.~Boulatov,
  %``A Model of three-dimensional lattice gravity,''
  Mod.\ Phys.\ Lett.\ A {\bf 7}, 1629 (1992)
  [hep-th/9202074].
  %%CITATION = HEP-TH/9202074;%%

%\cite{Freidel:2005qe}
\bibitem{Freidel:2005qe} 
  L.~Freidel,
  %``Group field theory: An Overview,''
  Int.\ J.\ Theor.\ Phys.\  {\bf 44}, 1769 (2005)
  [hep-th/0505016].
  %%CITATION = HEP-TH/0505016;%%

\bibitem{Oriti:2007qd} 
  D.~Oriti,
  %``Group field theory as the microscopic description of the quantum spacetime fluid: A New perspective on the continuum in quantum gravity,''
  PoS QG {\bf -PH}, 030 (2007)
  [arXiv:0710.3276 [gr-qc]].
  %%CITATION = ARXIV:0710.3276;%%

%\cite{Oriti:2011jm}
\bibitem{Oriti:2011jm} 
  D.~Oriti,
  %``The microscopic dynamics of quantum space as a group field theory,''
  arXiv:1110.5606 [hep-th].
  %%CITATION = ARXIV:1110.5606;%%

%\cite{Weingarten:1982mg}
\bibitem{Weingarten:1982mg} 
  D.~Weingarten,
  %``Euclidean Quantum Gravity On A Lattice,''
  Nucl.\ Phys.\ B {\bf 210}, 229 (1982);
  %%CITATION = NUPHA,B210,229;%%
%\cite{David:1984tx}
%\bibitem{David:1984tx} 
  F.~David,
  %``Planar Diagrams, Two-Dimensional Lattice Gravity and Surface Models,''
  Nucl.\ Phys.\ B {\bf 257}, 45 (1985);
  %%CITATION = NUPHA,B257,45;%%
%\cite{David:1985nj}
%\bibitem{David:1985nj} 
  %F.~David,
  %``A Model of Random Surfaces with Nontrivial Critical Behavior,''
  Nucl.\ Phys.\ B {\bf 257}, 543 (1985);
  %%CITATION = NUPHA,B257,543;%%
%\cite{Ambjorn:1985az}
%\bibitem{Ambjorn:1985az} 
  J.~Ambjorn, B.~Durhuus and J.~Frohlich,
  %``Diseases of Triangulated Random Surface Models, and Possible Cures,''
  Nucl.\ Phys.\ B {\bf 257}, 433 (1985);
  %%CITATION = NUPHA,B257,433;%%
%\cite{Kazakov:1985ea}
%\bibitem{Kazakov:1985ea} 
  V.~A.~Kazakov, A.~A.~Migdal and I.~K.~Kostov,
  %``Critical Properties of Randomly Triangulated Planar Random Surfaces,''
  Phys.\ Lett.\ B {\bf 157}, 295 (1985);
  %%CITATION = PHLTA,B157,295;%%
%\cite{Boulatov:1986mm}
%\bibitem{Boulatov:1986mm} 
  D.~V.~Boulatov, V.~A.~Kazakov, A.~A.~Migdal and I.~K.~Kostov,
  %``Possible Types Of Critical Behavior And The Mean Size Of Dynamically Triangulated Random Surfaces,''
  Phys.\ Lett.\ B {\bf 174}, 87 (1986);
  %%CITATION = PHLTA,B174,87;%%
%\cite{Boulatov:1986jd}
%\bibitem{Boulatov:1986jd} 
  %D.~V.~Boulatov, V.~A.~Kazakov, I.~K.~Kostov and A.~A.~Migdal,
  %``Analytical and Numerical Study of the Model of Dynamically Triangulated Random Surfaces,''
  Nucl.\ Phys.\ B {\bf 275}, 641 (1986).
  %%CITATION = NUPHA,B275,641;%%

%\cite{Di Francesco:1993nw}
\bibitem{Di Francesco:1993nw} 
  P.~Di Francesco, P.~H.~Ginsparg and J.~Zinn-Justin,
  %``2-D Gravity and random matrices,''
  Phys.\ Rept.\  {\bf 254}, 1 (1995)
  [hep-th/9306153].
  %%CITATION = HEP-TH/9306153;%%

%
\bibitem{Douglas:1989ve} 
  M.~R.~Douglas and S.~H.~Shenker,
  %``Strings in Less Than One-Dimension,''
  Nucl.\ Phys.\ B {\bf 335}, 635 (1990).
  %%CITATION = NUPHA,B335,635;%%

%\cite{Brezin:1990rb}
\bibitem{Brezin:1990rb} 
  E.~Brezin and V.~A.~Kazakov,
  %``Exactly Solvable Field Theories Of Closed Strings,''
  Phys.\ Lett.\ B {\bf 236}, 144 (1990).
  %%CITATION = PHLTA,B236,144;%%

%\cite{Gross:1989vs}
\bibitem{Gross:1989vs} 
  D.~J.~Gross and A.~A.~Migdal,
  %``Nonperturbative Two-Dimensional Quantum Gravity,''
  Phys.\ Rev.\ Lett.\  {\bf 64}, 127 (1990);
  %%CITATION = PRLTA,64,127;%%
%\cite{Gross:1989aw}
%\bibitem{Gross:1989aw} 
  %D.~J.~Gross and A.~A.~Migdal,
  %``A Nonperturbative Treatment Of Two-dimensional Quantum Gravity,''
  Nucl.\ Phys.\ B {\bf 340}, 333 (1990).
  %%CITATION = NUPHA,B340,333;%%

%\cite{Ginsparg:1993is}
\bibitem{Ginsparg:1993is} 
  P.~H.~Ginsparg and G.~W.~Moore,
  %``Lectures on 2-D gravity and 2-D string theory,''
  In *Boulder 1992, Proceedings, Recent directions in particle theory* 277-469. and Yale Univ. New Haven - YCTP-P23-92 (92,rec.Apr.93) 197 p. and Los Alamos Nat. Lab. - LA-UR-92-3479 (92,rec.Apr.93) 197 p
  [hep-th/9304011].
  %%CITATION = HEP-TH/9304011;%%



%\cite{Ambjorn:1994yv}
\bibitem{Ambjorn:1994yv} 
  J.~Ambjorn,
  %``Quantization of geometry,''
  hep-th/9411179.
  %%CITATION = HEP-TH/9411179;%%

%\cite{Marino:2004eq}
\bibitem{Marino:2004eq} 
  M.~Marino,
  %``Les Houches lectures on matrix models and topological strings,''
  hep-th/0410165.
  %%CITATION = HEP-TH/0410165;%%

%\cite{Brezin:1992yc}
\bibitem{Brezin:1992yc} 
  E.~Brezin and J.~Zinn-Justin,
  %``Renormalization group approach to matrix models,''
  Phys.\ Lett.\ B {\bf 288}, 54 (1992)
  [hep-th/9206035].
  %%CITATION = HEP-TH/9206035;%%


\bibitem{Alfaro:1992nq} 
  J.~Alfaro and P.~H.~Damgaard,
  %``The D = 1 matrix model and the renormalization group,''
  Phys.\ Lett.\ B {\bf 289}, 342 (1992)
  [hep-th/9206099].
  %%CITATION = HEP-TH/9206099;%%

%\cite{Higuchi:1993pu}
\bibitem{Higuchi:1993pu} 
  S.~Higuchi, C.~Itoi and N.~Sakai,
  %``Renormalization group approach to matrix models and vector models,''
  Prog.\ Theor.\ Phys.\ Suppl.\  {\bf 114}, 53 (1993)
  [hep-th/9307154].
  %%CITATION = HEP-TH/9307154;%%

%\cite{Higuchi:1994rv}
\bibitem{Higuchi:1994rv} 
  S.~Higuchi, C.~Itoi, S.~Nishigaki and N.~Sakai,
  %``Renormalization group flow in one and two matrix models,''
  Nucl.\ Phys.\ B {\bf 434}, 283 (1995)
  [Erratum-ibid.\ B {\bf 441}, 405 (1995)]
  [hep-th/9409009].
  %%CITATION = HEP-TH/9409009;%%

%\cite{Dasgupta:2003kk}
\bibitem{Dasgupta:2003kk} 
  S.~Dasgupta and T.~Dasgupta,
  %``Renormalization group approach to c = 1 matrix model on a circle and D-brane decay,''
  hep-th/0310106.
  %%CITATION = HEP-TH/0310106;%%

%
\bibitem{Ambjorn:1990ge} 
  J.~Ambjorn, B.~Durhuus and T.~Jonsson,
  %``Three-dimensional simplicial quantum gravity and generalized matrix models,''
  Mod.\ Phys.\ Lett.\ A {\bf 6}, 1133 (1991).
  %%CITATION = MPLAE,A6,1133;%%

%\cite{Sasakura:1990fs}
\bibitem{Sasakura:1990fs} 
  N.~Sasakura,
  %``Tensor model for gravity and orientability of manifold,''
  Mod.\ Phys.\ Lett.\ A {\bf 6}, 2613 (1991).
  %%CITATION = MPLAE,A6,2613;%%

%\cite{Gross:1991hx}
\bibitem{Gross:1991hx} 
  M.~Gross,
  %``Tensor models and simplicial quantum gravity in > 2-D,''
  Nucl.\ Phys.\ Proc.\ Suppl.\  {\bf 25A}, 144 (1992).
  %%CITATION = NUPHZ,25A,144;%%

%
%\cite{Gurau:2009tw}
\bibitem{Gurau:2009tw} 
  R.~Gurau,
  %``Colored Group Field Theory,''
  Commun.\ Math.\ Phys.\  {\bf 304}, 69 (2011)
  [arXiv:0907.2582 [hep-th]].
  %%CITATION = ARXIV:0907.2582;%%

\bibitem{Gurau:2010ba} 
  R.~Gurau,
  %``The 1/N expansion of colored tensor models,''
  Annales Henri Poincare {\bf 12}, 829 (2011)
  [arXiv:1011.2726 [gr-qc]].
  %%CITATION = ARXIV:1011.2726;%%

%\cite{Gurau:2011aq}
\bibitem{Gurau:2011aq} 
  R.~Gurau and V.~Rivasseau,
  %``The 1/N expansion of colored tensor models in arbitrary dimension,''
  Europhys.\ Lett.\  {\bf 95}, 50004 (2011)
  [arXiv:1101.4182 [gr-qc]].
  %%CITATION = ARXIV:1101.4182;%%

%\cite{Gurau:2011xq}
\bibitem{Gurau:2011xq} 
  R.~Gurau,
  %``The complete 1/N expansion of colored tensor models in arbitrary dimension,''
  Annales Henri Poincare {\bf 13}, 399 (2012)
  [arXiv:1102.5759 [gr-qc]].
  %%CITATION = ARXIV:1102.5759;%%

%\cite{Gurau:2011sk}
\bibitem{Gurau:2011sk} 
  R.~Gurau,
  %``The Double Scaling Limit in Arbitrary Dimensions: A Toy Model,''
  Phys.\ Rev.\ D {\bf 84}, 124051 (2011)
  [arXiv:1110.2460 [hep-th]].
  %%CITATION = ARXIV:1110.2460;%%


%\cite{Dartois:2013sra}
\bibitem{Dartois:2013sra} 
  S.~Dartois, R.~Gurau and V.~Rivasseau,
  %``Double Scaling in Tensor Models with a Quartic Interaction,''
  arXiv:1307.5281 [hep-th].
  %%CITATION = ARXIV:1307.5281;%%



%\cite{Gurau:2010nd}
\bibitem{Gurau:2010nd} 
  R.~Gurau,
  %``Lost in Translation: Topological Singularities in Group Field Theory,''
  Class.\ Quant.\ Grav.\  {\bf 27}, 235023 (2010)
  [arXiv:1006.0714 [hep-th]].
  %%CITATION = ARXIV:1006.0714;%%

%
%\cite{Ayala:1993fj}
\bibitem{Ayala:1993fj} 
  C.~Ayala,
  %``Renormalization group approach to matrix models in two-dimensional quantum gravity,''
  Phys.\ Lett.\ B {\bf 311}, 55 (1993)
  [hep-th/9304090].
  %%CITATION = HEP-TH/9304090;%%






\bibitem{Wetterich:1993yh}
C.~Wetterich,
%`Exact evolution equation for the effective potential,''
Phys.\ Lett.\ B {\bf 301}, 90 (1993).
%%CITATION = PHLTA,B301,90;%%

\bibitem{Berges:2000ew}
  J.~Berges, N.~Tetradis and C.~Wetterich,
 %  ``Non-perturbative renormalization flow in quantum field
%theory and
 %  statistical physics,''
  %
  Phys.\ Rept.\  {\bf 363} (2002) 223
  [hep-ph/0005122].
  %%CITATION = HEP-PH 0005122;%%
%
%\cite{Polonyi:2001se}
\bibitem{Polonyi:2001se}
  J.~Polonyi,
  %``Lectures on the functional renormalization group
%method,''
  Central Eur.\ J.\ Phys.\  {\bf 1} (2003) 1. 
 [hep-th/0110026].
  %%CITATION = HEP-TH 0110026;%%
%
%\cite{Pawlowski:2005xe}
\bibitem{Pawlowski:2005xe}
  J.~M.~Pawlowski,
 % ``Aspects of the functional renormalisation group,''
  Annals Phys.\  {\bf 322} (2007) 2831 
  [arXiv:hep-th/0512261].
  %%CITATION = APNYA,322,2831;%%
%

%\cite{Gies:2006wv}
\bibitem{Gies:2006wv}
  H.~Gies, in {\it Renormalization Group and Effective Field Theory Approaches to Many-Body Systems,
Lecture Notes in Physics, Vol 852}, pp 287-348 (2012)
  %``Introduction to the functional RG and applications to gauge theories,''
  arXiv:hep-ph/0611146. %.
  %%CITATION = HEP-PH/0611146;%%

%\cite{Delamotte:2007pf}
\bibitem{Delamotte:2007pf} 
  B.~Delamotte,
  %``An Introduction to the nonperturbative renormalization group,''
  Lect.\ Notes Phys.\  {\bf 852}, 49 (2012)
  [cond-mat/0702365 [COND-MAT]].
  %%CITATION = COND-MAT/0702365;%%
  
%\cite{Rosten:2010vm}
\bibitem{Rosten:2010vm}
  O.~J.~Rosten,
%  ``Fundamentals of the Exact Renormalization Group,''
  arXiv:1003.1366 [hep-th].
  %%CITATION = ARXIV:1003.1366;%%

\bibitem{metzner2011} 
W.~Metzner et al., Rev. Mod. Phys. {\bf 84}, 299 (2012).

%\cite{Braun:2011pp}
\bibitem{Braun:2011pp} 
  J.~Braun,
  %``Fermion Interactions and Universal Behavior in Strongly Interacting Theories,''
  J.\ Phys.\ G {\bf 39}, 033001 (2012)
  [arXiv:1108.4449 [hep-ph]].
  %%CITATION = ARXIV:1108.4449;%%

%\cite{Sfondrini:2010zm}
\bibitem{Sfondrini:2010zm} 
  A.~Sfondrini and T.~A.~Koslowski,
  %``Functional Renormalization of Noncommutative Scalar Field Theory,''
  Int.\ J.\ Mod.\ Phys.\ A {\bf 26}, 4009 (2011)
  [arXiv:1006.5145 [hep-th]].
  %%CITATION = ARXIV:1006.5145;%%

%\cite{Grosse:2004yu}
\bibitem{Grosse:2004yu} 
  H.~Grosse and R.~Wulkenhaar,
  %``Renormalization of phi**4 theory on noncommutative R**4 in the matrix base,''
  Commun.\ Math.\ Phys.\  {\bf 256}, 305 (2005)
  [hep-th/0401128].
  %%CITATION = HEP-TH/0401128;%%

%\cite{Disertori:2006nq}
\bibitem{Disertori:2006nq} 
  M.~Disertori, R.~Gurau, J.~Magnen and V.~Rivasseau,
  %``Vanishing of Beta Function of Non Commutative Phi**4(4) Theory to all orders,''
  Phys.\ Lett.\ B {\bf 649}, 95 (2007)
  [hep-th/0612251].
  %%CITATION = HEP-TH/0612251;%%


%\cite{Litim:2001up}
\bibitem{Litim:2001up} 
  D.~F.~Litim,
  %``Optimized renormalization group flows,''
  Phys.\ Rev.\ D {\bf 64}, 105007 (2001)
  [hep-th/0103195].
  %%CITATION = HEP-TH/0103195;%%

\bibitem{Canet:2003qd} 
  L.~Canet, B.~Delamotte, D.~Mouhanna and J.~Vidal,
  %``Nonperturbative renormalization group approach to the Ising model: A Derivative expansion at order partial**4,''
  Phys.\ Rev.\ B {\bf 68}, 064421 (2003)
  [hep-th/0302227].
  %%CITATION = HEP-TH/0302227;%%

%\cite{Litim:2010tt}
\bibitem{Litim:2010tt} 
  D.~F.~Litim and D.~Zappala,
  %``Ising exponents from the functional renormalisation group,''
  Phys.\ Rev.\ D {\bf 83}, 085009 (2011)
  [arXiv:1009.1948 [hep-th]].
  %%CITATION = ARXIV:1009.1948;%%

%\cite{Gies:2002af}
\bibitem{Gies:2002af} 
  H.~Gies,
  %``Running coupling in Yang-Mills theory: A flow equation study,''
  Phys.\ Rev.\ D {\bf 66}, 025006 (2002)
  [hep-th/0202207].
  %%CITATION = HEP-TH/0202207;%%

%\cite{Benedetti:2012dx}
\bibitem{Benedetti:2012dx} 
  D.~Benedetti and F.~Caravelli,
  %``The Local potential approximation in quantum gravity,''
  JHEP {\bf 1206}, 017 (2012)
  [Erratum-ibid.\  {\bf 1210}, 157 (2012)]
  [arXiv:1204.3541 [hep-th]].
  %%CITATION = ARXIV:1204.3541;%%

%\cite{Dietz:2012ic}
\bibitem{Dietz:2012ic} 
  J.~A.~Dietz and T.~R.~Morris,
  %``Asymptotic safety in the f(R) approximation,''
  JHEP {\bf 1301}, 108 (2013)
  [arXiv:1211.0955 [hep-th]].
  %%CITATION = ARXIV:1211.0955;%%

%
\bibitem{Litim:1998wk} 
  D.~F.~Litim and J.~M.~Pawlowski,
  %``On gauge invariance and Ward identities for the Wilsonian renormalization group,''
  Nucl.\ Phys.\ Proc.\ Suppl.\  {\bf 74}, 325 (1999)
  [hep-th/9809020].
  %%CITATION = HEP-TH/9809020;%%

%\cite{Litim:1998nf}
\bibitem{Litim:1998nf} 
  D.~F.~Litim and J.~M.~Pawlowski,
  %``On gauge invariant Wilsonian flows,''
  hep-th/9901063.
  %%CITATION = HEP-TH/9901063;%%



%
\bibitem{BenGeloun:2012pu} 
  J.~Ben Geloun and D.~O.~Samary,
  %``3D Tensor Field Theory: Renormalization and One-loop $\beta$-functions,''
  Annales Henri Poincare {\bf 14}, 1599 (2013)
  [arXiv:1201.0176 [hep-th]].
  %%CITATION = ARXIV:1201.0176;%%

%\cite{BenGeloun:2012yk}
\bibitem{BenGeloun:2012yk} 
  J.~Ben Geloun,
  %``Two and four-loop $\beta$-functions of rank 4 renormalizable tensor field theories,''
  Class.\ Quant.\ Grav.\  {\bf 29}, 235011 (2012)
  [arXiv:1205.5513 [hep-th]].
  %%CITATION = ARXIV:1205.5513;%%

%\cite{Geloun:2012qn}
\bibitem{Geloun:2012qn} 
  J.~Ben Geloun,
  %``Asymptotic Freedom of Rank 4 Tensor Group Field Theory,''
  arXiv:1210.5490 [hep-th].
  %%CITATION = ARXIV:1210.5490;%%

%\cite{Weinberg:1980gg}
\bibitem{Weinberg:1980gg}
  S.~Weinberg,
%  ``Ultraviolet Divergences In Quantum Theories Of Gravitation,''
%\href{http://www.slac.stanford.edu/spires/find/hep/www?irn=784877}{SPIRES entry}
{\it  In *Hawking, S.W., Israel, W.: General Relativity*, 790-831}
(Cambridge University Press, Cambridge, 1980).

\bibitem{Gies:2009hq} 
  H.~Gies and M.~M.~Scherer,
  %``Asymptotic safety of simple Yukawa systems,''
  Eur.\ Phys.\ J.\ C {\bf 66}, 387 (2010)
  [arXiv:0901.2459 [hep-th]].
  %%CITATION = ARXIV:0901.2459;%%

%\cite{Gies:2013pma}
\bibitem{Gies:2013pma} 
  H.~Gies, S.~Rechenberger, M.~M.~Scherer and L.~Zambelli,
  %``An asymptotic safety scenario for gauged chiral Higgs-Yukawa models,''
  arXiv:1306.6508 [hep-th].
  %%CITATION = ARXIV:1306.6508;%%

%
\bibitem{Reuter:1996cp} 
  M.~Reuter,
  %``Nonperturbative evolution equation for quantum gravity,''
  Phys.\ Rev.\ D {\bf 57}, 971 (1998)
  [hep-th/9605030].

%\cite{Reuter:2012id}
\bibitem{Reuter:2012id} 
  M.~Reuter and F.~Saueressig,
  %``Quantum Einstein Gravity,''
  New J.\ Phys.\  {\bf 14}, 055022 (2012)
  [arXiv:1202.2274 [hep-th]].
  %%CITATION = ARXIV:1202.2274;%%

%






\end{thebibliography}
\end{document}